\documentclass[preprintnumbers,10pt,nofootinbib]{revtex4}

\usepackage{amsmath,latexsym,amssymb,amsfonts}
\usepackage[dvips]{graphicx,color}
\usepackage{bm}

\usepackage{xcolor}
\usepackage{soul}

\begin{document}

\title{\textbf{Euclidean wormholes in Gauss-Bonnet-dilaton gravity}}
\author{
\textsc{Xiao Yan Chew$^{a,b}$}\footnote{{\tt xychew998@gmail.com}},
\textsc{Gansukh Tumurtushaa$^{c, d}$}\footnote{{\tt gansuh.mgl@gmail.com}}
and
\textsc{Dong-han Yeom$^{a,b}$}\footnote{{\tt innocent.yeom{}@{}gmail.com}}
}

\affiliation{
$^{a}$\small{Department of Physics Education, Pusan National University, Busan 46241, Republic of Korea}\\
$^{b}$\small{Research Center for Dielectric and Advanced Matter Physics, Pusan National University, Busan 46241, Republic of Korea}\\
$^{c}$\small{Department of Science Education, Jeju National University, Jeju 63243, Republic of Korea}\\
$^{d}$\small{Center for Theoretical Physics of the Universe, Institute for Basic Science, Daejeon 34051, Republic of Korea}
}

\begin{abstract}
We investigate Euclidean wormholes in Gauss-Bonnet-dilaton gravity to explain the creation of the universe from nothing. We considered two types of dilaton couplings (i.e., the string-inspired model and the Gaussian model) and we obtained qualitatively similar results. There can exist Euclidean wormholes that explain the possible origin of our universe, where the dilaton field is located over the barrier of dilaton potential. This solution can exist even if dilaton potential does not satisfy slow-roll conditions. In addition, the probability is higher than that of the Hawking-Moss instanton with the same final condition. Therefore, Euclidean wormholes in Gauss-Bonnet-dilaton gravity are a possible and probable scenario, which explains the origin of our universe.
\end{abstract}

\maketitle

\newpage

\tableofcontents


\section{Introduction}

To understand the origin of the universe, we need to investigate two topics, i.e., (1) the fundamental action of quantum gravity that explains our universe and (2) the way to evaluate the wave function of the universe for a given action. Regarding the first topic, string theory provides a clue; specifically, the Einstein gravity needs to be modified owing to higher order stringy corrections \cite{Metsaev:1987zx}. 

Within the context of modifying general relativity, the Gauss-Bonnet term has played a major part in the past few decades as the low-energy effective theory of the ultimate quantum gravity. Because the Gauss-Bonnet term is a geometric invariant, i.e., by itself, it is a total derivative in four-dimensions (4D), it does not contribute to gravitational dynamics. Thus, to account for its effects, one can introduce additional degrees of freedom, such as a scalar field (i.e., dilaton) coupling $\xi(\phi)$, and couple it to the Gauss-Bonnet term. However, Ref.~\cite{Glavan:2019inb} has recently proposed a novel 4D Einstein-Gauss-Bonnet gravity, in which the authors considered a coupling constant $\alpha$ instead of dilaton coupling $\xi(\phi)$ functions. Using an unusual action principle and rescaling of the coupling constant, $\alpha\rightarrow\alpha/(D-4)$, the nontrivial contribution of gravitational dynamics is realized in $D$ dimensional spacetime. Here, the factor $1/(D-4)$ regularizes the otherwise vanishing contribution from the Gauss-Bonnet term, which allows the $D\rightarrow4$  limit at the level of equations of motion~\cite{Glavan:2019inb}. Therefore, it was possible for the topological Gauss-Bonnet invariant to have a finite local dynamics in $4D$ spacetime with some symmetries. There has been an ongoing discussion regarding the nature and definiteness of the four-dimensional limit of the Einstein-Gauss-Bonnet theory~\cite{Lu:2020iav}.

In this paper, let us focus on the leading term of the modified gravity action via string theory that can be well categorized by the Gauss-Bonnet-dilaton gravity \cite{Kanti:1995vq}:
\begin{eqnarray}\label{eq:action}
S = \int d^{4}x \sqrt{-g} \left[ \frac{R}{2\kappa^{2}} - \frac{1}{2} \left(\nabla \phi\right)^{2} - V(\phi) + \frac{1}{2} \xi(\phi) R_{\mathrm{GB}}^{2} \right],
\end{eqnarray}
where $R$ is the Ricci scalar, $\kappa^{2} = 8\pi$, $\phi$ is the dilaton field with potential $V(\phi)$, and $R_{\mathrm{GB}}^{2}$ is the Gauss-Bonnet term. If we extend this model, we may generalize the form of the coupling term $\xi(\phi)$ between the dilaton field and the Gauss-Bonnet term as well as the potential $V(\phi)$ of the dilaton field $\phi$.

In relation to Eq.~(\ref{eq:action}), the cosmological application of models with the Gauss-Bonnet term in the 4D Friedmann universe has been previously studied in many works, e.g., inflation~\cite{Koh:2014bka, Guo:2009uk} and dark energy~\cite{Calcagni:2005im} (and references therein). The gravitational wave constraints on the coupling strength of the Gauss-Bonnet term have been discussed in Refs.~\cite{Gong:2017kim}.  A notable feature of such models in the abovementioned literature~\cite{Koh:2014bka, Guo:2009uk, Calcagni:2005im} is that the Gauss-Bonnet term affects the cosmological dynamics only through the coupling function $\xi(\phi)$. Thus, in this study, we focus on a case in which the Gauss-Bonnet term is non-minimally coupled to the dilaton field via $\xi(\phi)$ rather than having a coupling constant.

Using this theoretical background, we can investigate the wave function of the universe \cite{DeWitt:1967yk}. There have been several proposals. However, in this study, we focus on the Euclidean path-integral approach after the Wick-rotation to the Euclidean time $\tau = it$ \cite{Hartle:1983ai}. The Euclidean path-integral is defined from the in-state $(h_{ab}^{\mathrm{in}}, \phi^{\mathrm{in}})$ to out-state $(h_{ab}^{\mathrm{out}}, \phi^{\mathrm{out}})$, where $h_{ab}$ is the three-metric, and $\phi$ is the field value at a given slice:
\begin{eqnarray}
\langle h_{ab}^{\mathrm{out}}, \phi^{\mathrm{out}} | h_{ab}^{\mathrm{in}}, \phi^{\mathrm{in}} \rangle = \int \mathcal{D}g_{\mu\nu} \mathcal{D}\phi \; e^{-S_{\mathrm{E}}},
\end{eqnarray}
where we sum over all geometries that connect from the in-state to the out-state, and $S_{\mathrm{E}}$ is the Euclidean action. This path-integral will be approximated well using the steepest-descent approximation, i.e., the path-integral can be approximated by summing over on-shell histories, or so-called instantons. The instanton solutions should connect from the in-state to out-state \cite{Chen:2018aij}. If the instanton solution for the in-state and the solution for the out-state are disconnected on the Euclidean section, we can independently define the wave function only for the out-state; in this case, there is no initial boundary; hence, this proposal is also known as the no-boundary proposal \cite{Hartle:1983ai}.

The no-boundary proposal requires that for a given out-state, we sum over all regular and compact Euclidean instantons. In the Einstein gravity, typically, the solution looks like a Hawking-Moss instanton \cite{Hawking:1981fz}. However, if we include the Gauss-Bonnet-dilaton gravity terms, we can identify new solutions, e.g., Euclidean wormholes \cite{Chen:2016ask}. In our previous paper \cite{Tumurtushaa:2018agq}, we investigated whether there exists a Euclidean wormhole solution that could explain the creation of a wormhole \textit{ex nihilo} from quantum gravitational fluctuations. However, this is not a unique application. If we consider a homogeneous analytic continuation \cite{Lee:2012qv}, we can apply this instanton to quantum cosmology and the initial state of the universe \cite{Hwang:2011mp}. We can ask whether the Gauss-Bonnet-dilaton gravity can provide new instantons for the no-boundary proposal, which can compete with typical Hawking-Moss type solutions, e.g., \cite{Chen:2019cmw}.

This paper is organized as follows. In Sec.~\ref{sec:mod}, we describe the details of the Gauss-Bonnet-dilaton gravity model. In Sec.~\ref{sec:res}, we show several numerical instanton solutions as well as their probabilities. Finally, in Sec.~\ref{sec:con}, we summarize this paper and comment on possible future applications.

\section{\label{sec:mod}Model}

The action of the Gauss-Bonnet-dilaton gravity is \cite{Kanti:1995vq}
\begin{eqnarray}
S = \int d^{4}x \sqrt{-g} \left[ \frac{R}{2\kappa^{2}} - \frac{1}{2} \left(\nabla \phi\right)^{2} - V(\phi) + \frac{1}{2} \xi(\phi) R_{\mathrm{GB}}^{2} \right],
\end{eqnarray}
where $R$ is the Ricci scalar, $\kappa^{2} = 8\pi$, $\phi$ is the dilaton field with potential $V(\phi)$, and
\begin{eqnarray}
R_{\mathrm{GB}}^{2} = R_{\mu\nu\rho\sigma}R^{\mu\nu\rho\sigma} - 4 R_{\mu\nu}R^{\mu\nu} + R^{2}
\end{eqnarray}
is the Gauss-Bonnet term. The conventional choice of the coupling function $\xi(\phi)$ following the type-II superstring theory is \cite{Metsaev:1987zx}
\begin{eqnarray}
\xi_{1}(\phi) = \lambda e^{- c \phi},
\end{eqnarray}
where $\lambda$ and $c$ are the model-dependent parameters. However, recently, there have been some investigations for the Gaussian model
\begin{eqnarray}
\xi_{2}(\phi) = \lambda e^{- c \phi^{2}},
\end{eqnarray}
where $\lambda$ and $c$ are the model-dependent parameters. To reveal the genuine properties of the Gauss-Bonnet-dilaton gravity model, it is worthwhile to study a non-conventional choice of the dilaton coupling function.

\subsection{Equations of motion}

In this theoretical background, we study instantons with the following $O(4)$-symmetric metric ansatz:
\begin{eqnarray}
ds_{\mathrm{E}}^{2} = d\tau^{2} + a^{2}(\tau) d\Omega_{3}^{2},
\end{eqnarray}
where $\tau$ is the Euclidean time, $a(\tau)$ is the metric function for the scale factor, and
\begin{eqnarray}
d\Omega_{3}^{2} = d\chi^{2} + \sin^{2}{\chi} \left( d\theta^{2} + \sin^{2}{\theta} d\varphi^{2} \right)
\end{eqnarray}
is the three-sphere.

Then, the equations of motion in Euclidean signatures are \cite{Cai:2008ht,Koh:2014bka}
\begin{eqnarray}
\label{Eq1} H^{2} &=& \frac{\kappa^{2}}{3} \left[ \frac{1}{2} \dot{\phi}^{2} - V + \frac{3}{\kappa^{2} a^{2}} - 12 \dot{\xi} H  \left( - H^{2} + \frac{1}{a^{2}} \right) \right],\\
\label{Eq2} \dot{H} &=& -\frac{\kappa^{2}}{2} \left[ \dot{\phi}^{2} + \frac{2}{\kappa^{2} a^{2}} + 4\ddot{\xi} \left(-H^{2} + \frac{1}{a^{2}} \right) + 4 \dot{\xi} H \left(-2 \dot{H} + H^{2} - \frac{3}{a^{2}} \right) \right],\\
\label{Eq3} 0 &=& \ddot{\phi} + 3 H \dot{\phi} - V' - 12 \xi' \left( - H^{2} + \frac{1}{a^{2}} \right) \left( \dot{H} + H^{2} \right),
\end{eqnarray}
where $H \equiv \dot{a}/a$. Eqs.~(\ref{Eq2}) and (\ref{Eq3}) will be used to numerically solve the variables. Eq.~(\ref{Eq1}) will be the constraint equation, where it is simplified (if $a \neq 0$) to
\begin{eqnarray}
0 = 6a \left( 1 - \dot{a}^{2} \right) - 24 \kappa^{2} \dot{a} \left( 1 - \dot{a}^{2} \right) \dot{\phi} \xi' + \kappa^{2} a^{3} \left( \dot{\phi}^{2} - 2V \right),
\end{eqnarray}
where $'$ denotes the derivation with respect to $\phi$, while $\dot{}$ denotes the derivation with respect to the Euclidean time $\tau$.

We present Eqs.~(\ref{Eq2}) and (\ref{Eq3}) for $\ddot{a}$ and $\ddot{\phi}$ \cite{Tumurtushaa:2018agq}:
\begin{eqnarray}
\label{eq:a}\ddot{a} &=& - \frac{a^{2}}{2} \mathcal{F},\\
\label{eq:phi}\ddot{\phi} &=& V' - 3 \frac{\dot{a}}{a} \dot{\phi} - \frac{6 \xi' \left(1 - \dot{a}^{2}\right)}{a} \mathcal{F},
\end{eqnarray}
where
\begin{equation}
\mathcal{F} \equiv \frac{2a (1-\dot{a}^{2}) + \kappa^{2} a^{3} \dot{\phi}^{2} - 4\kappa^{2}\xi''(- a \dot{\phi}^{2} + a \dot{a}^{2} \dot{\phi}^{2} ) - 4\kappa^{2} \xi'( - a V' + a \dot{a}^{2} V' + 6 \dot{a} \dot{\phi} - 6 \dot{a}^{3} \dot{\phi})}{a^{4} - 4\kappa^{2} a^{3} \dot{a} \dot{\phi} \xi' + 24 \kappa^{2} \xi'^{2} - 48 \kappa^{2} \dot{a}^{2} \xi'^{2} + 24 \kappa^{2} \dot{a}^{4} \xi'^{2}}.
\end{equation}

In general, to solve a compact and regular instanton solution with a generic potential, the problem becomes a boundary value problem; hence, one needs to tune the initial condition to satisfy boundary conditions. Because this procedure is technically complicated, for simplicity, we solve the solution in a different way; first, we fix the form of $\phi$, then, solve $V$ by reverse engineering \cite{Kanno:2012zf}. This is a solution searching technique; in realistic examples, we need to fix the potential first and solve the field value later. If we accept this technical approach, we are allowed to use the following form of the field:
\begin{eqnarray}
\phi(\tau) - \phi_{0} = \frac{\left(\phi_{1} - \phi_{0} \right)}{12\pi} \left[ 12 \frac{\tau}{\Delta} - 8\sin 2\frac{\tau}{\Delta} + \sin 4 \frac{\tau}{\Delta} \right]
\end{eqnarray}
for $0 \leq \tau \leq \pi\Delta$, while $\phi(\tau) = \phi_{1}$ for $\tau > \pi\Delta$. Here, $\phi_{0}$, $\phi_{1}$, and $\Delta$ are free parameters. The equation for $V$ now becomes $\dot{V} = \dot{\phi} V'$, where
\begin{equation}\label{eq:V}
V' = \frac{a^{5} \ddot{\phi} + 6 \kappa^{2} a^{3} (1 - 3 \dot{a}^{2}) \dot{\phi}^{2} \xi' - 12 \xi' (1 - \dot{a}^{2})^{2} ( 6 \kappa^{2} \dot{a} \dot{\phi} \xi' - a - 2 a \kappa^{2} \dot{\phi}^{2} \xi'' - 2 a \kappa^{2} \ddot{\phi} \xi') + a^{4} \dot{a} \dot{\phi} (3 - 4 \kappa^{2} \ddot{\phi} \xi')}{a^{4} (a - 4\kappa^{2} \dot{a} \dot{\phi} \xi')}.
\end{equation}
Therefore, finally, we will solve Eqns.~(\ref{eq:a}) and (\ref{eq:V}) for $a(\tau)$ and $V(\tau)$.

\begin{figure}
\begin{center}
\includegraphics[scale=0.5]{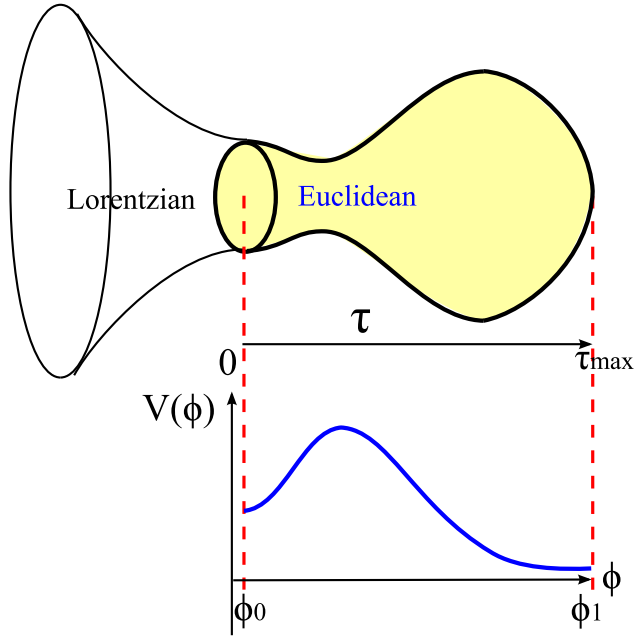}
\caption{\label{fig:fig}Homogeneous analytic continuations of a Euclidean wormhole.}
\end{center}
\end{figure}

\subsection{Initial conditions}

We impose the following initial conditions at $\tau = 0$:
\begin{eqnarray}
a(0) &=& \left[ \frac{6}{\kappa^{2} \left(2 V_{0} - \dot{\phi}^{2}(0) \right)} \right]^{1/2},\\
\dot{a}(0) &=& 0,\\
V(0) &=& V_{0},
\end{eqnarray}
where, for convenience, we choose
\begin{eqnarray}
\phi(0) &=& \phi_{0} = 0,\\
\dot{\phi}(0) &=& 0.
\end{eqnarray}
From this initial condition, we solve the solution up to $\tau = \tau_{\mathrm{max}}$, where
\begin{eqnarray}
a(\tau_{\mathrm{max}}) &=& 0,\\
\dot{a} (\tau_{\mathrm{max}}) &=& -1
\end{eqnarray}
are required for the regular end.

\begin{figure}
\begin{center}
\includegraphics[scale=0.3,angle=-90]{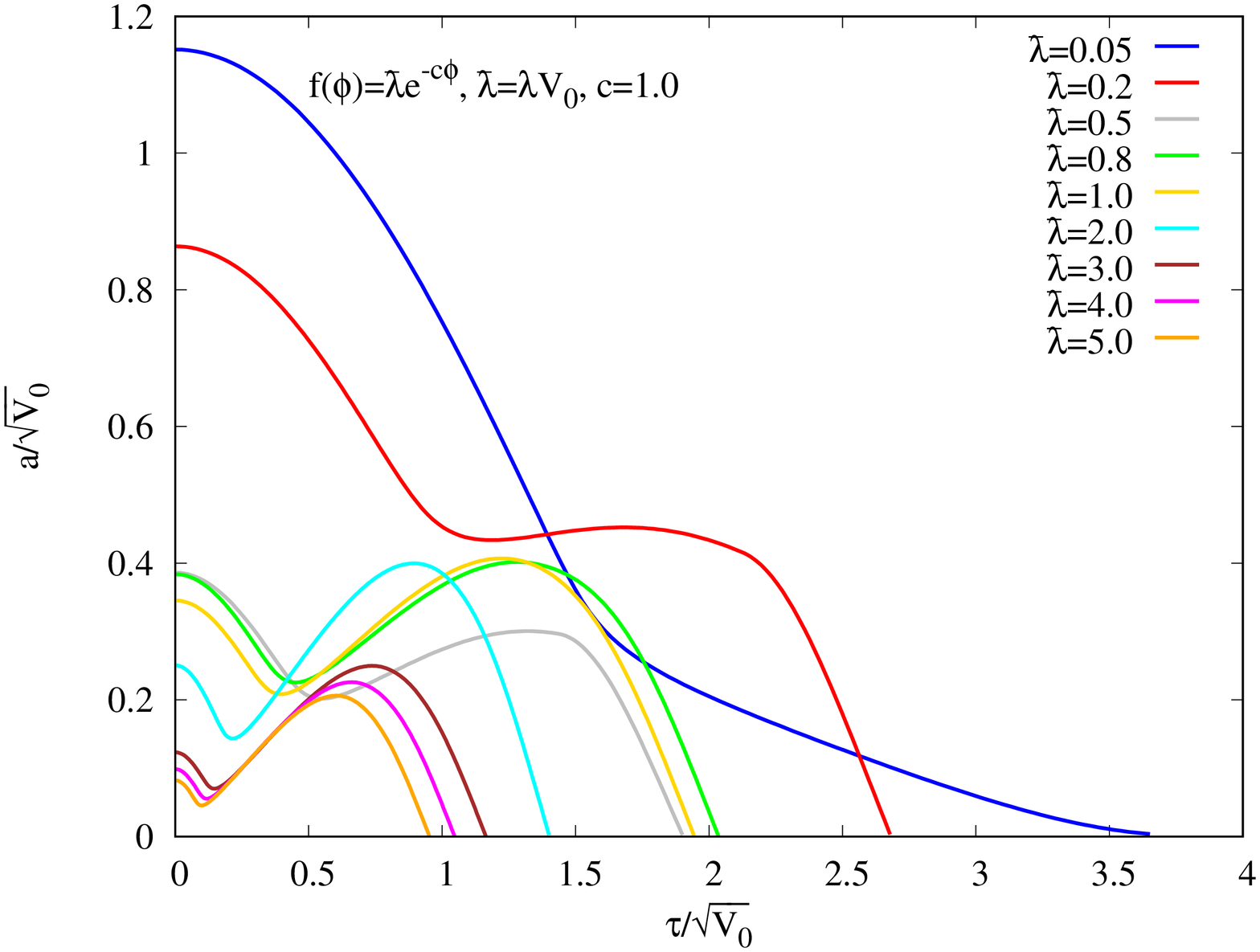}
\includegraphics[scale=0.3,angle=-90]{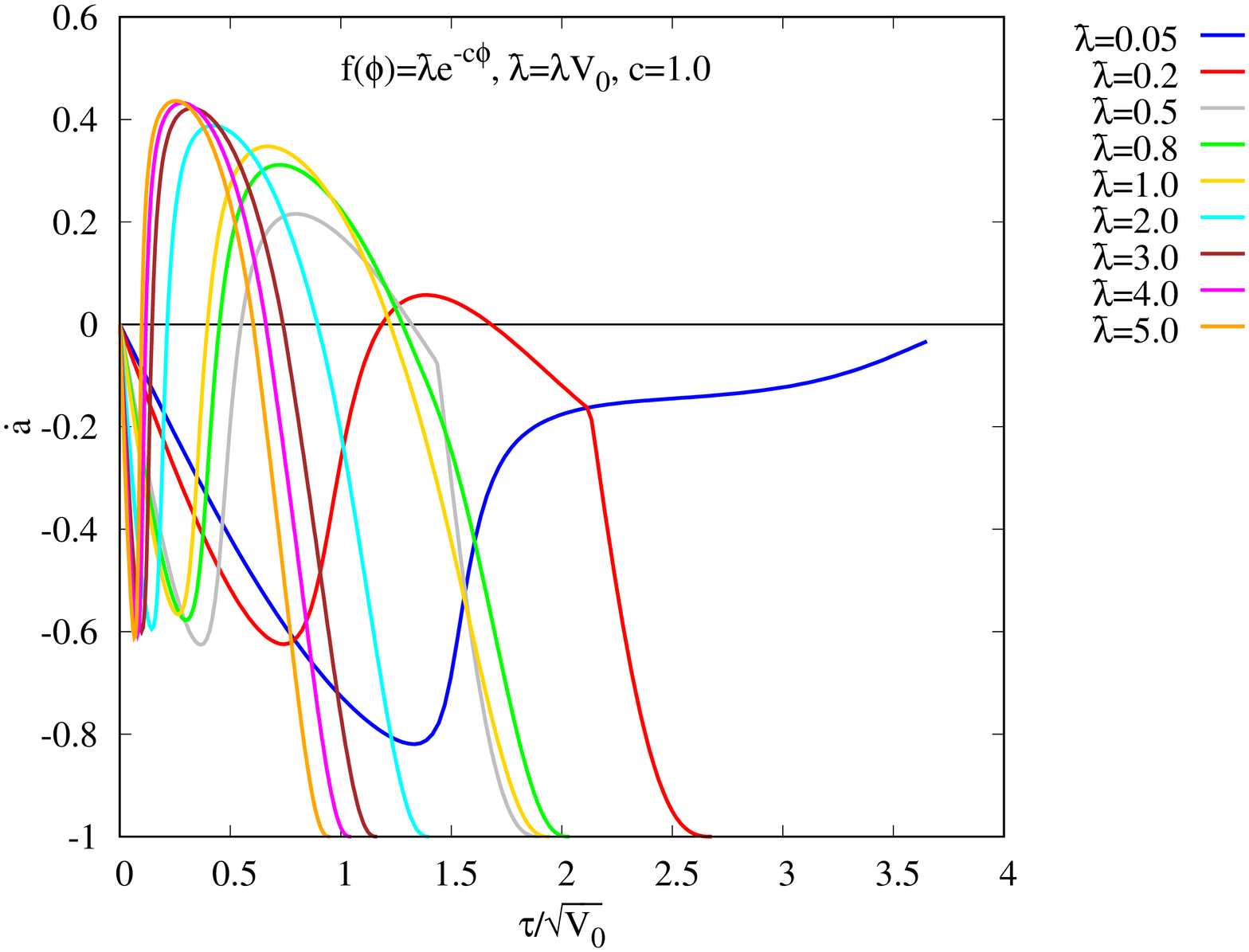}
\includegraphics[scale=0.3,angle=-90]{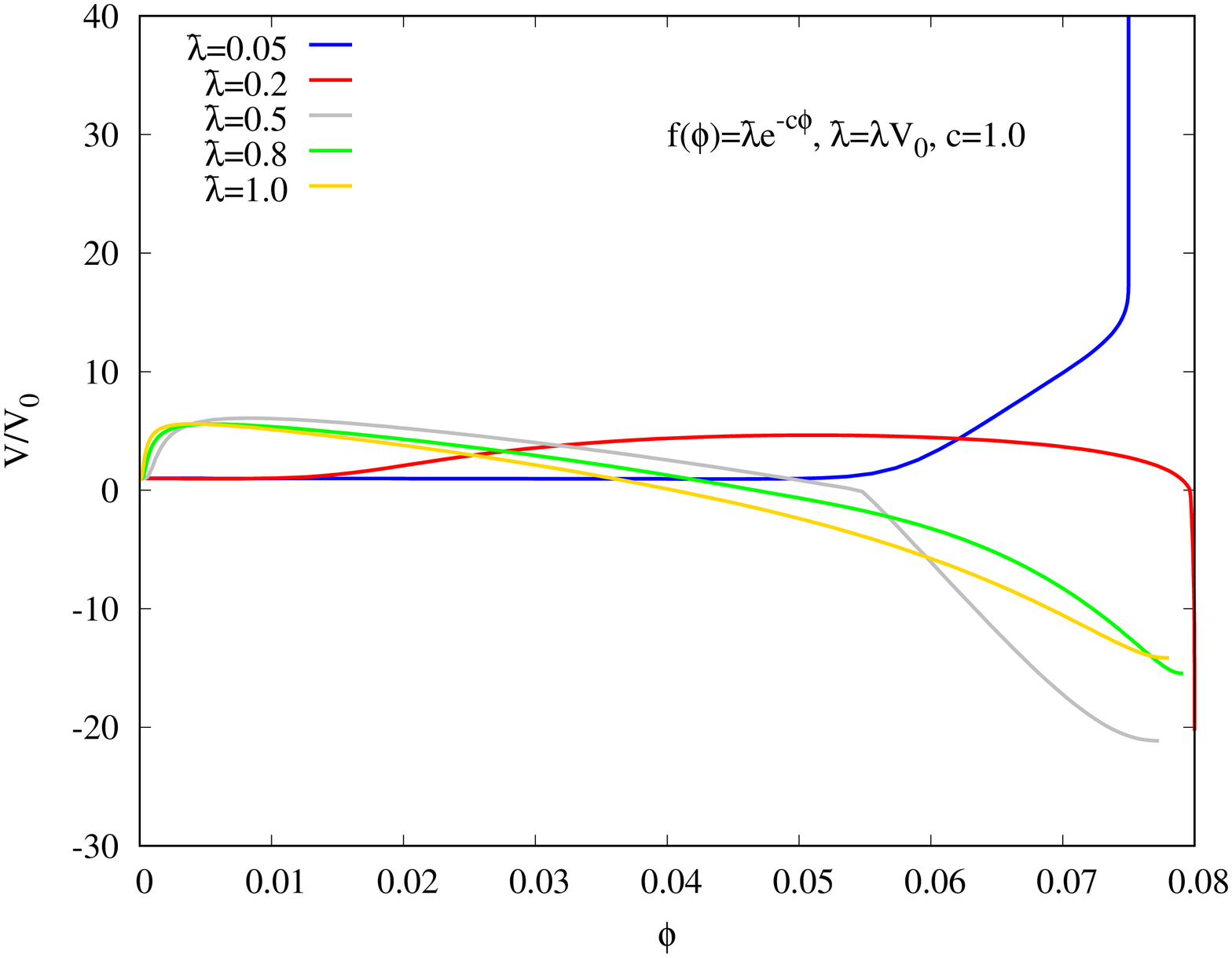}
\includegraphics[scale=0.3,angle=-90]{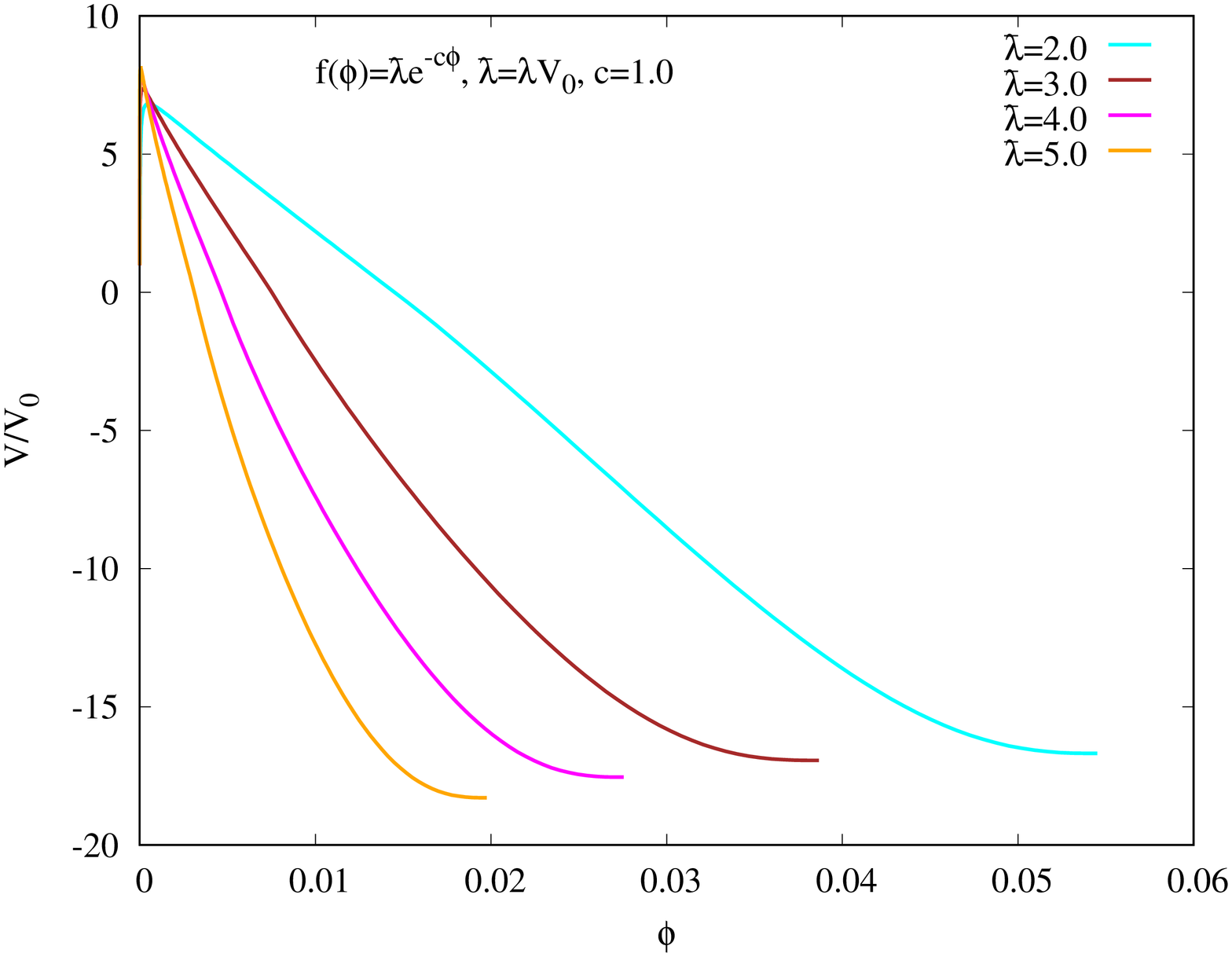}
\caption{\label{fig:cc1}$a/\sqrt{V_{0}}$ (upper left) and $\dot{a}$ (upper right) versus $\tau/\sqrt{V_{0}}$, and $V(\phi)/V_{0}$ (lower left and right) for $\xi_{1}$, where $c = 1$, $\phi_{0} = 0$, $\Delta = 0.8$, and varying $\bar{\lambda} \equiv \lambda V_{0}$; to satisfy the boundary condition, we need to tune $\phi_{1}$ and $V_{0}$.}
\end{center}
\end{figure}
\begin{figure}
\begin{center}
\includegraphics[scale=0.3,angle=-90]{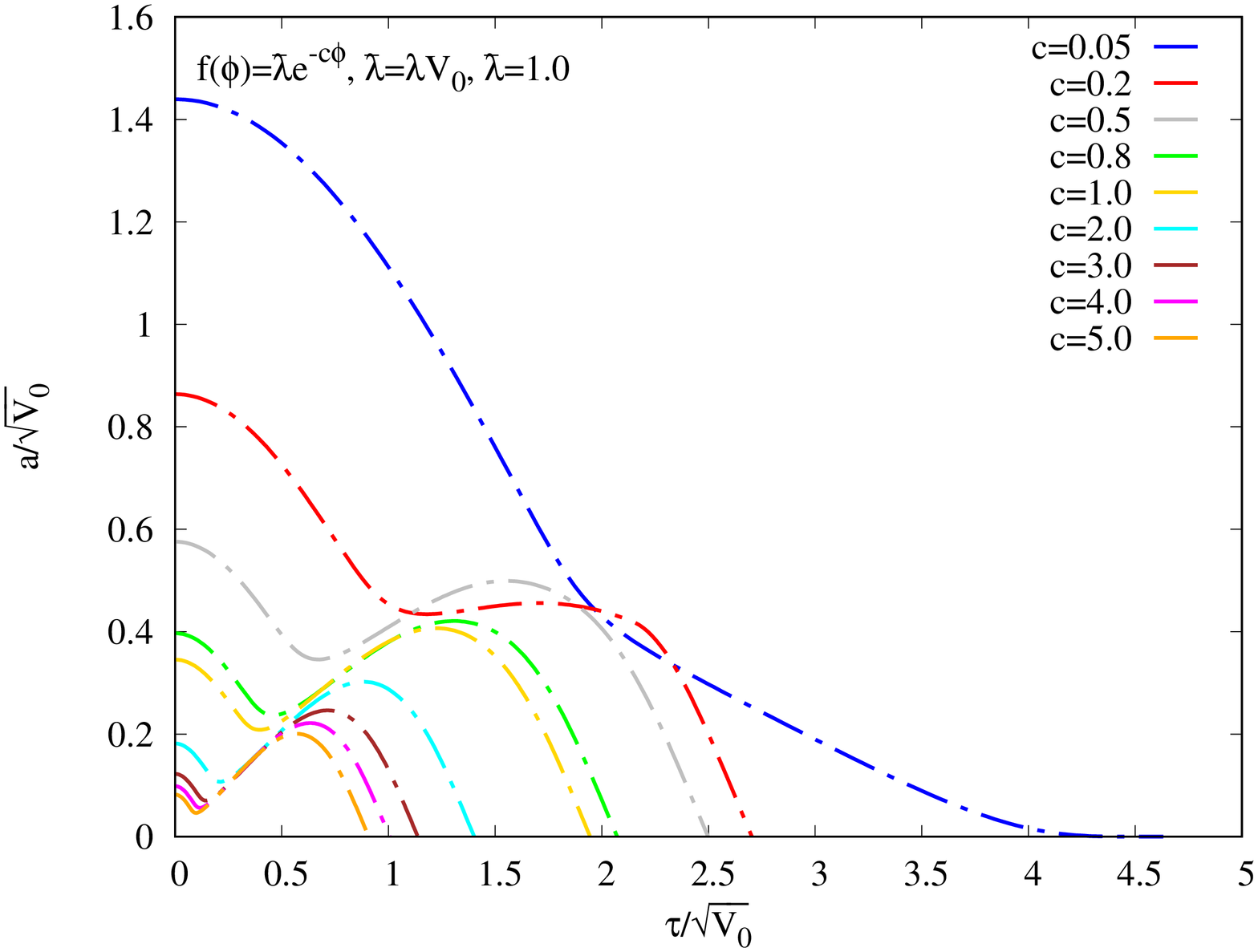}
\includegraphics[scale=0.3,angle=-90]{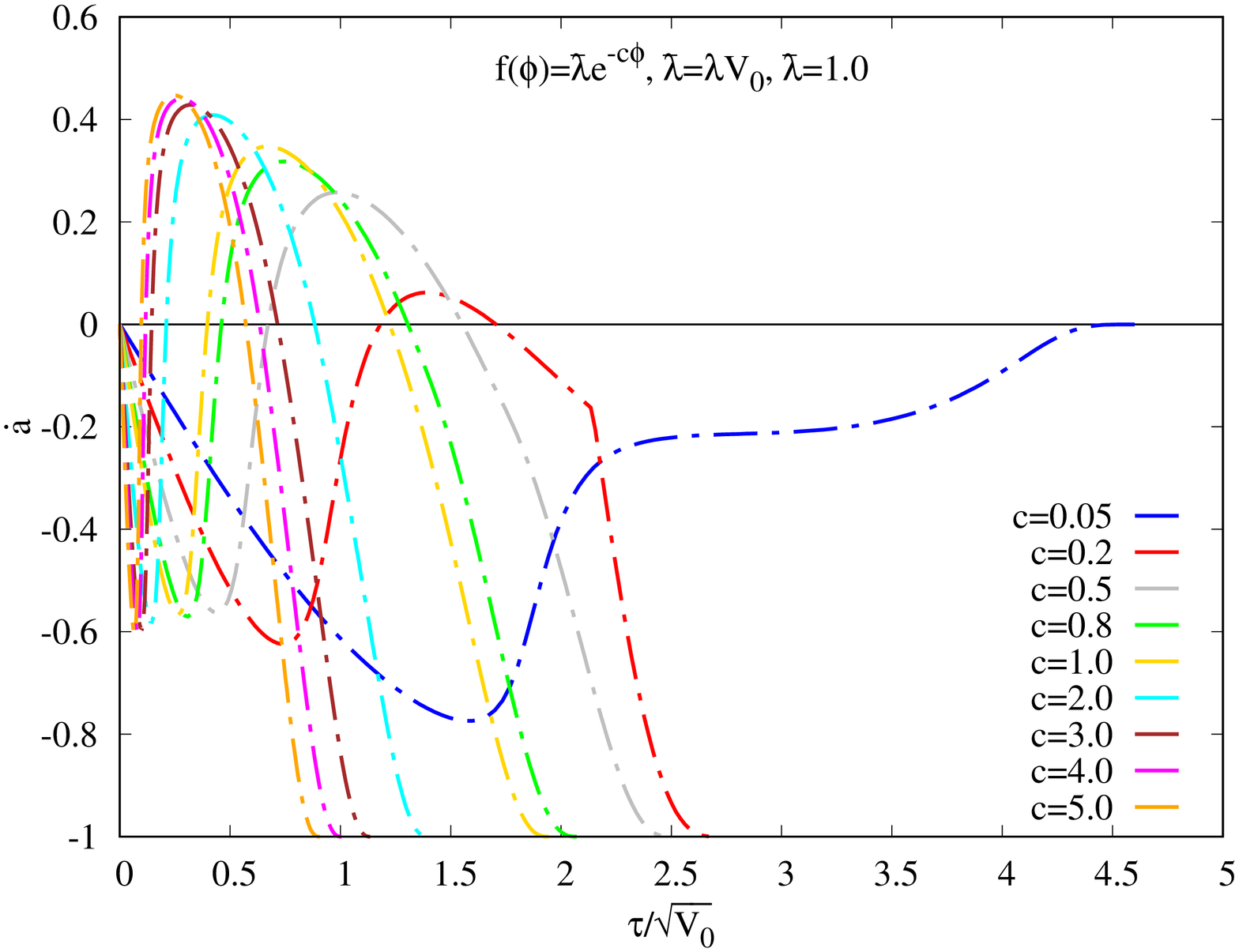}
\includegraphics[scale=0.3,angle=-90]{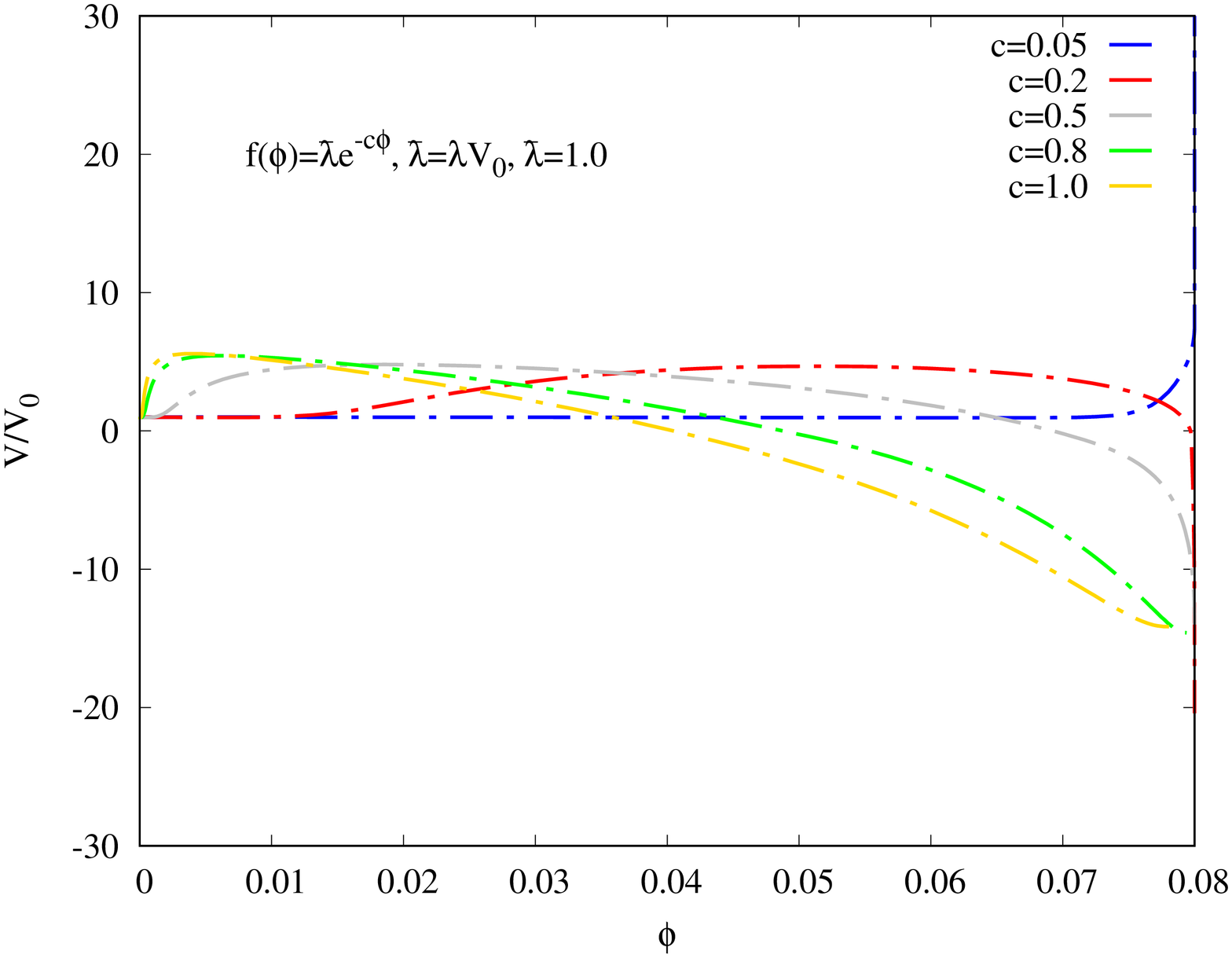}
\includegraphics[scale=0.3,angle=-90]{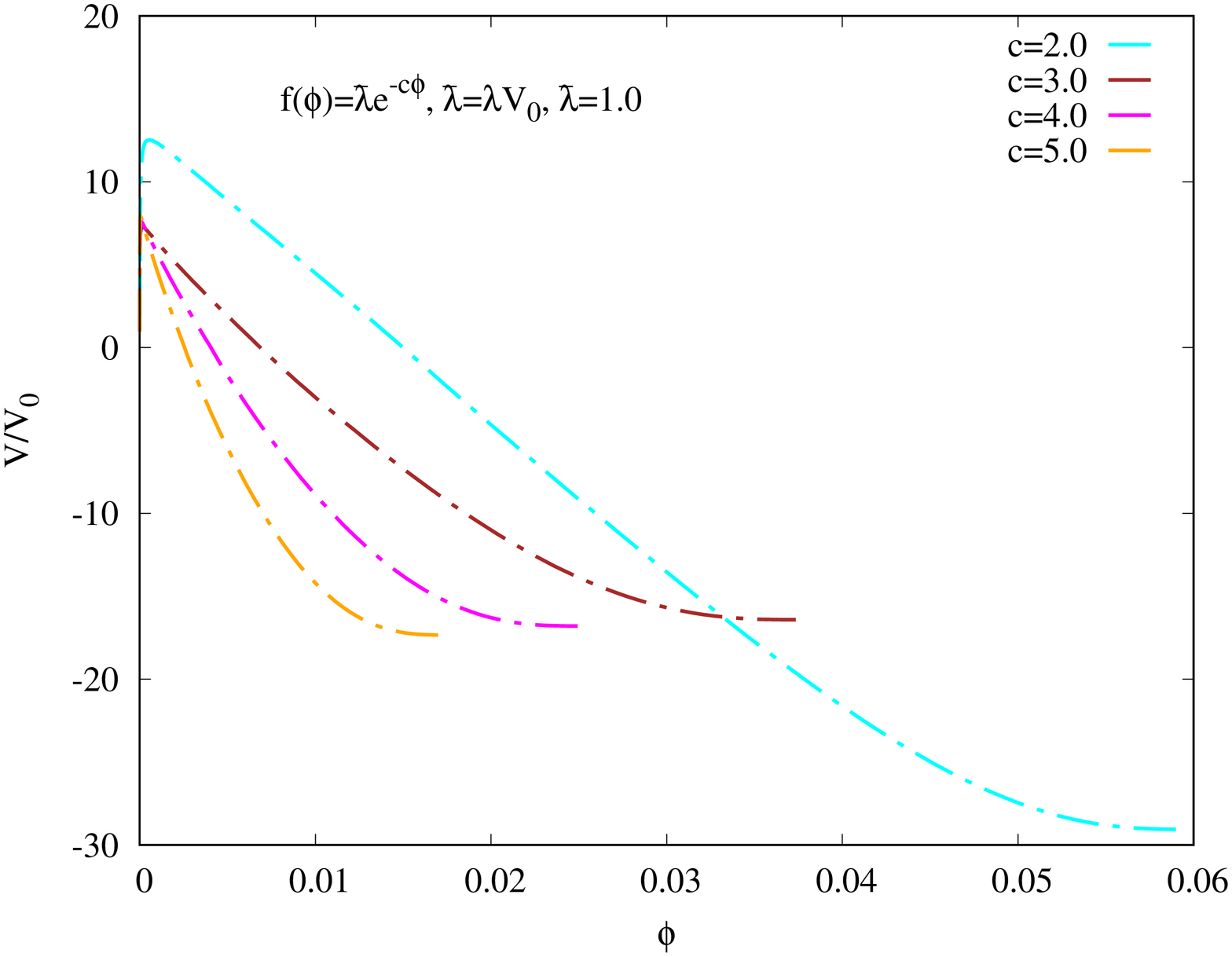}
\caption{\label{fig:lambda1}$a/\sqrt{V_{0}}$ (upper left) and $\dot{a}$ (upper right) versus $\tau/\sqrt{V_{0}}$, and $V(\phi)/V_{0}$ (lower left and right) for $\xi_{1}$, where $\lambda V_{0} = 1$, $\phi_{0} = 0$, $\Delta = 0.8$, and varying $c$; to satisfy the boundary condition, we need to tune $\phi_{1}$ and $V_{0}$.}
\end{center}
\end{figure}

\subsection{Homogeneous analytic continuations}

In this study, we are interested in constructing a Euclidean wormhole solution with the homogeneous analytic continuation. First, to have a Euclidean wormhole solution, we need to check what is the necessary condition. Geometrically, we need $\ddot{a} > 0$ when $\dot{a} = 0$ and $a > 0$. If we plug these conditions into the equations of motion, then necessary conditions are
\begin{eqnarray}
\ddot{a} &=& - \frac{a^{3}}{2} \left( \frac{2 + \kappa^{2} \dot{\phi}^{2} (a^{2} + 4 \xi'') + 4 \kappa^{2} \xi' V'}{a^{4} + 24 \kappa^{2} \xi'^{2}} \right),\\
\dot{\phi}^{2} &=& - \frac{6}{\kappa^{2} a^{2}} + 2 V.
\end{eqnarray}
These equations imply that the necessary conditions for the Euclidean wormhole become
\begin{eqnarray}
\frac{1}{2\kappa^{2}} + \xi' V' + \left( \frac{a^{2}}{4} + \xi'' \right) \dot{\phi}^{2} &<& 0,\\
V - \frac{3}{\kappa^{2} a^{2}} &\geq& 0.
\end{eqnarray}
These conditions are useful to determine the properties of wormholes. This means that the Euclidean wormhole throat can exist only if $V > 0$ and $\xi' V' < 0$ are satisfied.

Even before we obtain numerical solutions, we can expect that the solution will look like Fig.~\ref{fig:fig}. In the Euclidean domain, the field varies from $\phi_{0}$ to $\phi_{1}$ as the Euclidean time varies from $0$ to $\tau_{\mathrm{max}}$. If there is a Euclidean wormhole in the Euclidean domain, the geometry will look like a yellow-colored region.

After we obtain a solution, we introduce the homogeneous analytic continuation $t = -i \tau$ at $\tau = 0$. Note that at this point, we already imposed the condition of $\dot{a} = 0$ and $\dot{\phi} = 0$. Hence, after the Wick-rotation, the reality of the metric and the matter field are naturally guaranteed. After the Wick-rotation, the universe will evolve from the initial condition $\phi = \phi_{0}$. Unlike the inhomogeneous analytic continuation, the bottleneck of the Euclidean section will not be naked to the Lorentzian geometry. However, because the volume of the yellow-colored region of Fig.~\ref{fig:fig} is larger than that of the trivial Hawking-Moss instanton, the probability will be different. In some sense, one can say that there is an enhancement of the probability owing the existence of the potential barrier if we consider the Gauss-Bonnet-dilaton gravity. In the next section, we will confirm these expectations in detail with numerical solutions.

\subsection{Scaling behavior}

It is worthwhile to show the following scaling dependence of the solution. If we rescale the parameters, or if we introduce the conformal transformation,
\begin{eqnarray}
a &\rightarrow& \frac{a}{\sqrt{V_{0}}},\\
d\tau &\rightarrow& \frac{d\tau}{\sqrt{V_{0}}},
\end{eqnarray}
then
\begin{eqnarray}
S_{\mathrm{E}} \rightarrow - \frac{1}{V_{0}} \int d^{4}x \sqrt{g} \left[ \frac{R}{2\kappa^{2}} - \frac{1}{2} \left(\nabla \phi\right)^{2} - \frac{V(\phi)}{V_{0}} + \frac{1}{2} V_{0} \xi(\phi) R_{\mathrm{GB}}^{2} \right].
\end{eqnarray}
Therefore, from this scaling, the dynamics (equations of motion) is invariant even though we change $V_{0}$ and $\lambda \rightarrow \lambda V_{0}$ at the same time. Thus, when we interpret the potential, we will use this freedom to choose $V_{0}$. However, as we vary $V_{0}$, the physical probability must be scaled: $e^{-S_{\mathrm{E}}} \rightarrow e^{-S_{\mathrm{E}}/V_{0}}$.

\begin{figure}
\begin{center}
\includegraphics[scale=0.3,angle=-90]{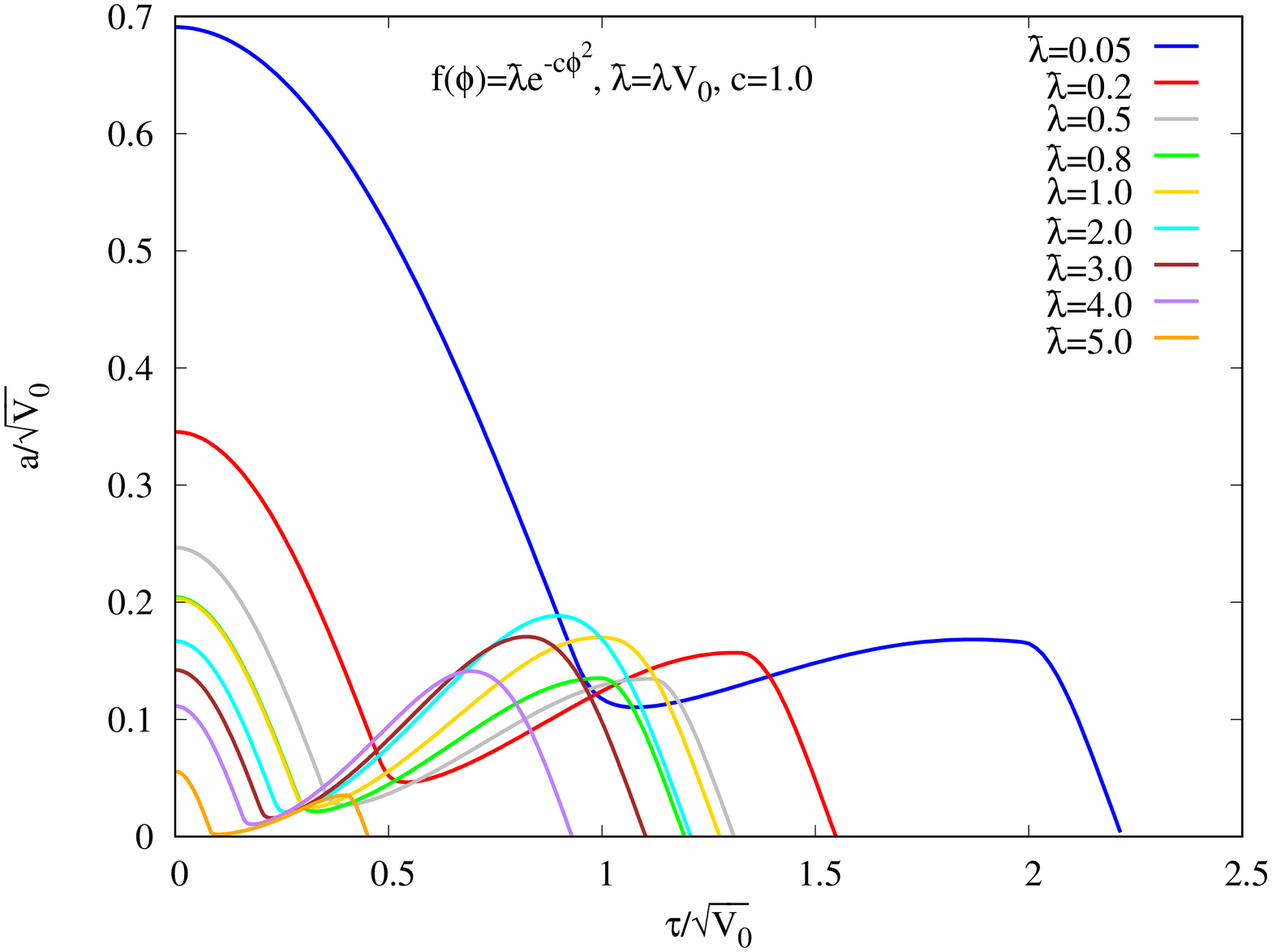}
\includegraphics[scale=0.3,angle=-90]{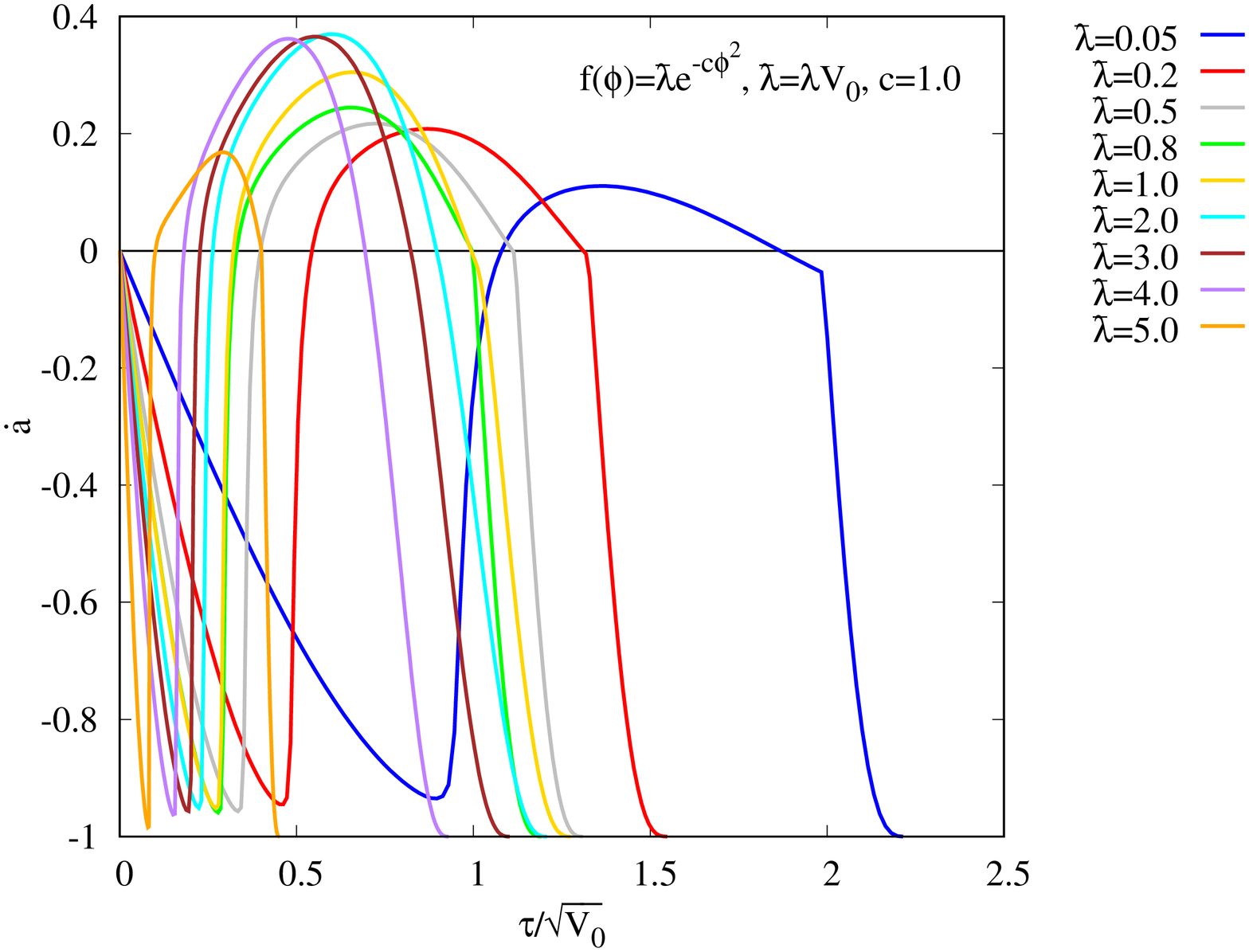}
\includegraphics[scale=0.3,angle=-90]{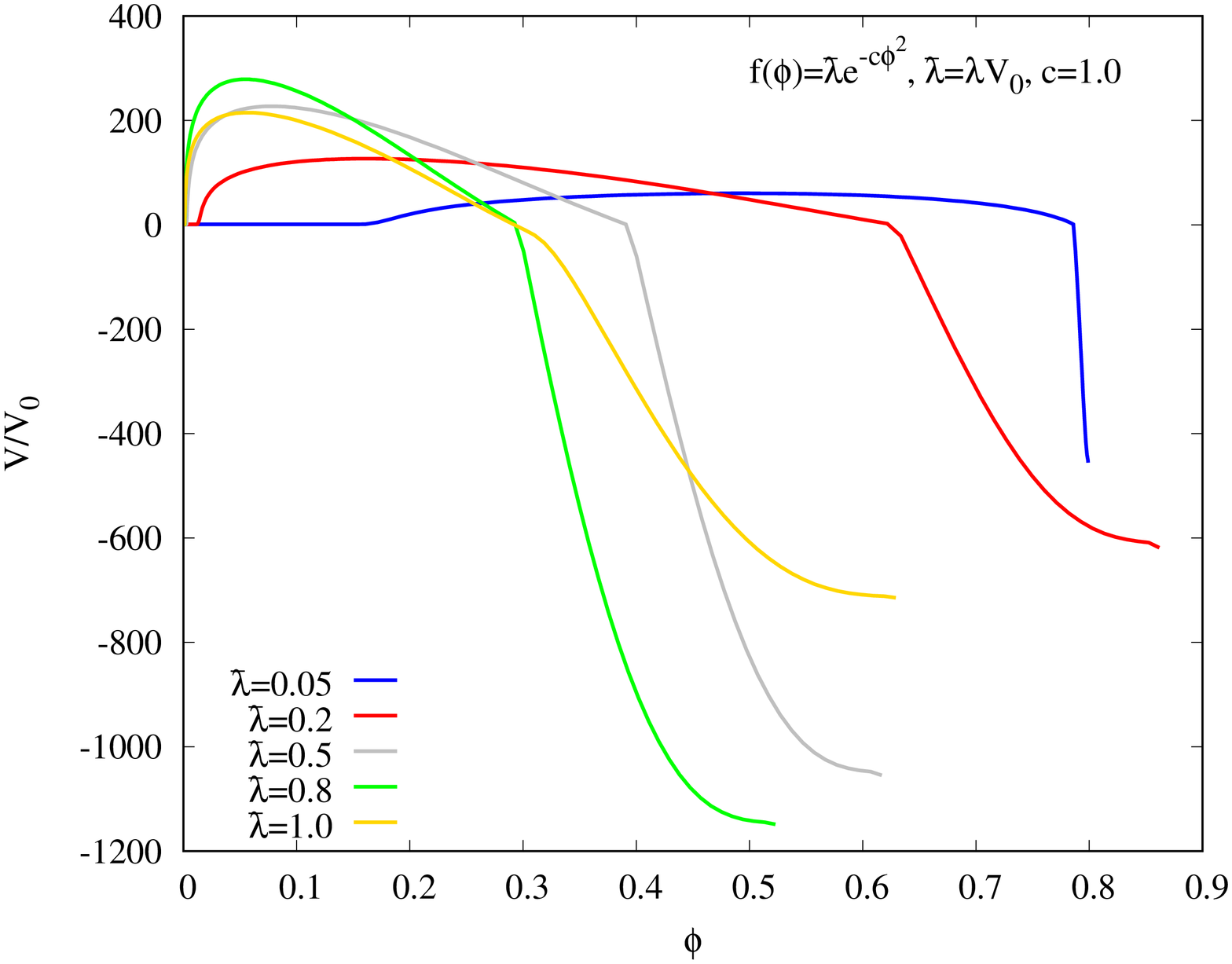}
\includegraphics[scale=0.3,angle=-90]{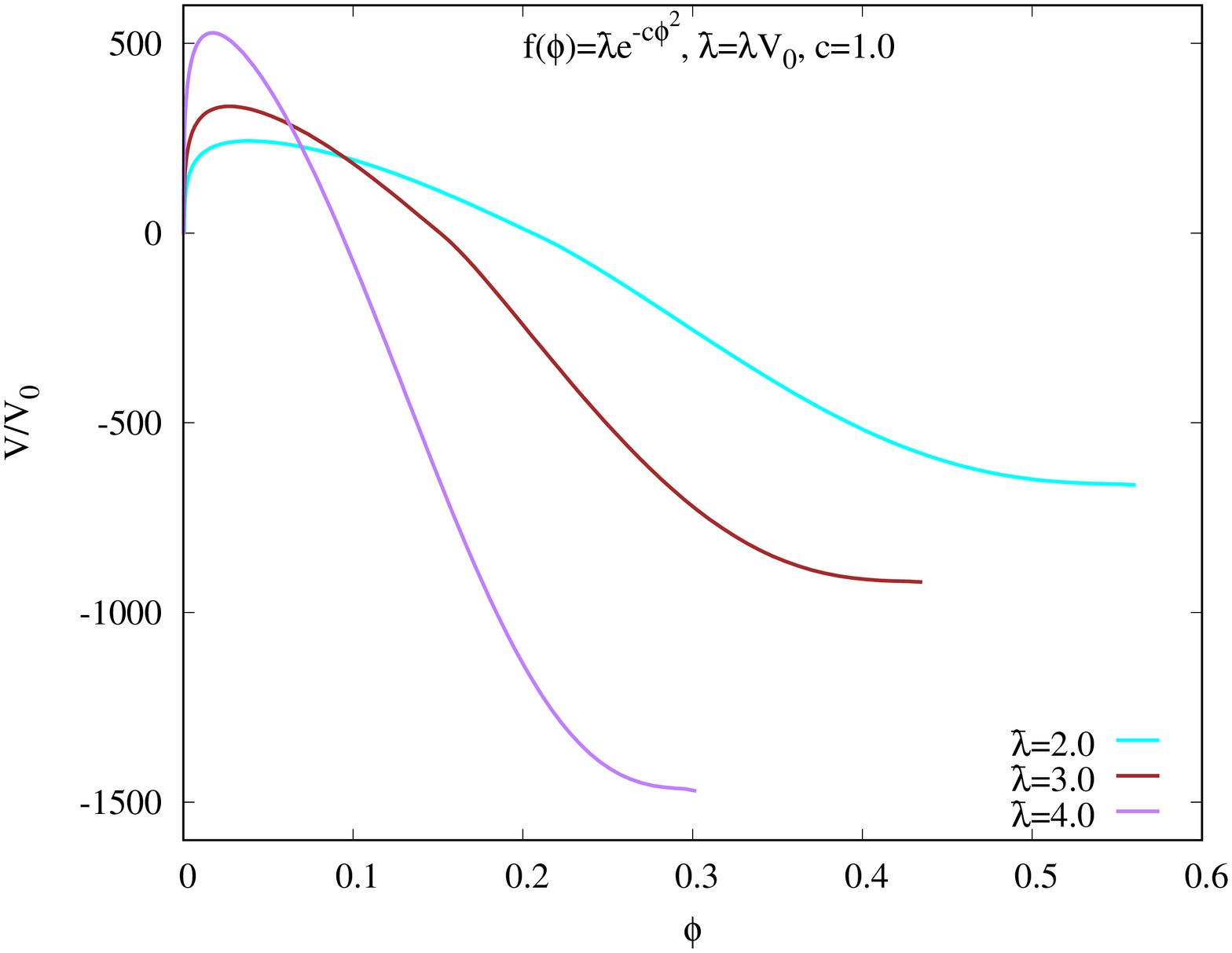}
\caption{\label{fig:cc2}$a/\sqrt{V_{0}}$ (upper left) and $\dot{a}$ (upper right) versus $\tau/\sqrt{V_{0}}$, and $V(\phi)/V_{0}$ (lower left and right) for $\xi_{2}$, where $c = 1$, $\phi_{0} = 0$, $\Delta = 0.8$, and varying $\bar{\lambda} \equiv \lambda V_{0}$; to satisfy the boundary condition, we need to tune $\phi_{1}$ and $V_{0}$.}
\end{center}
\end{figure}
\begin{figure}
\begin{center}
\includegraphics[scale=0.3,angle=-90]{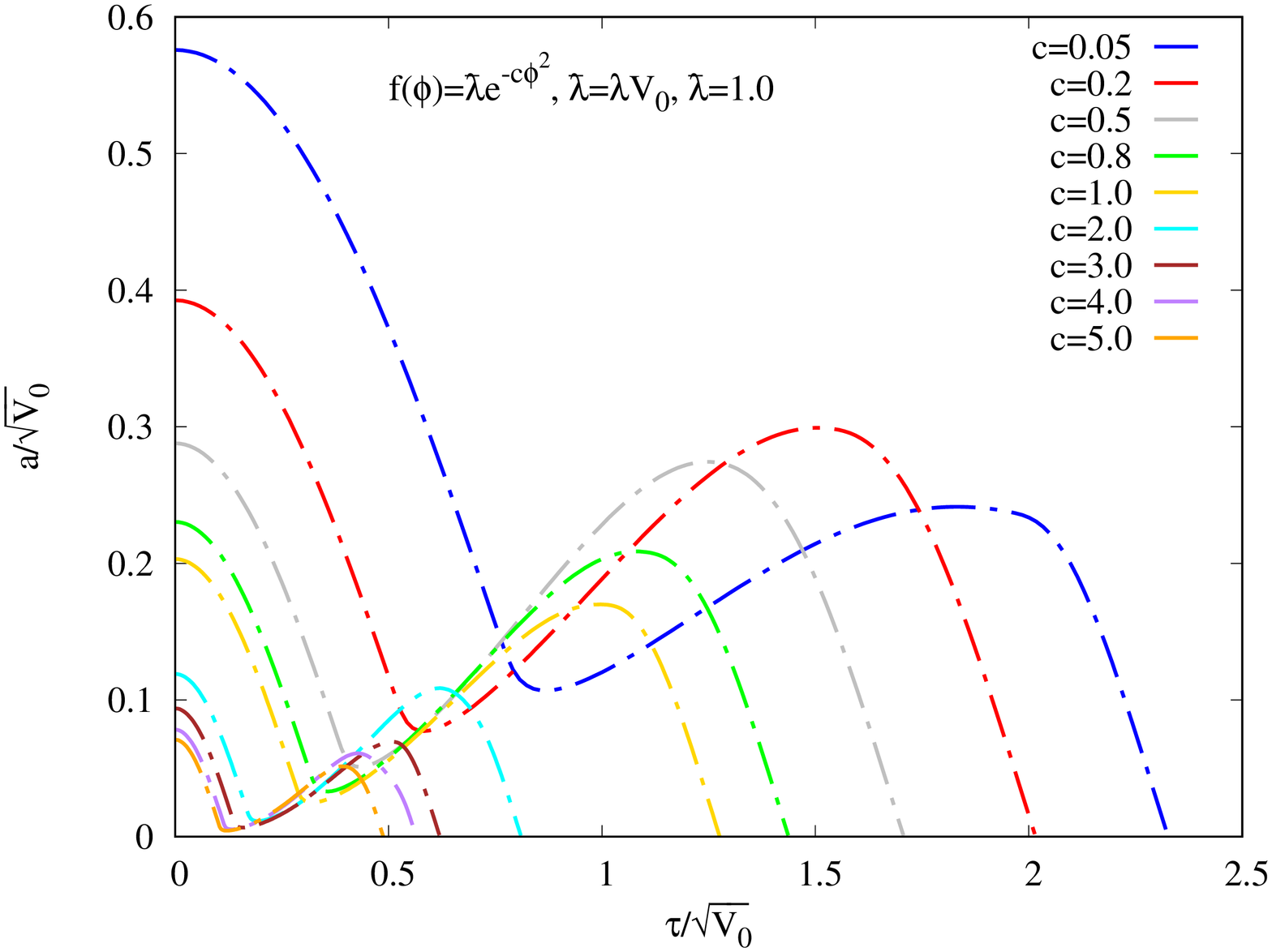}
\includegraphics[scale=0.3,angle=-90]{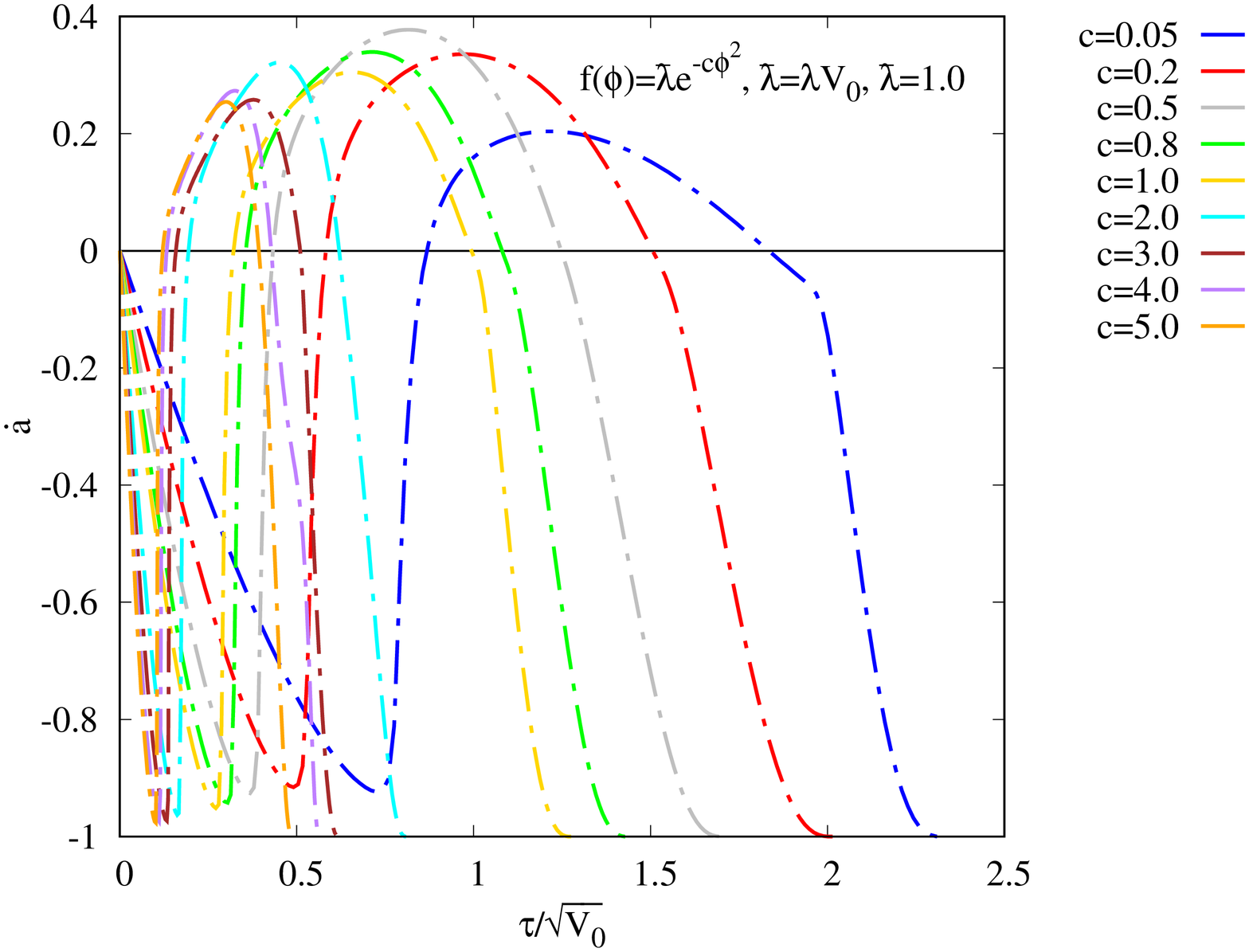}
\includegraphics[scale=0.3,angle=-90]{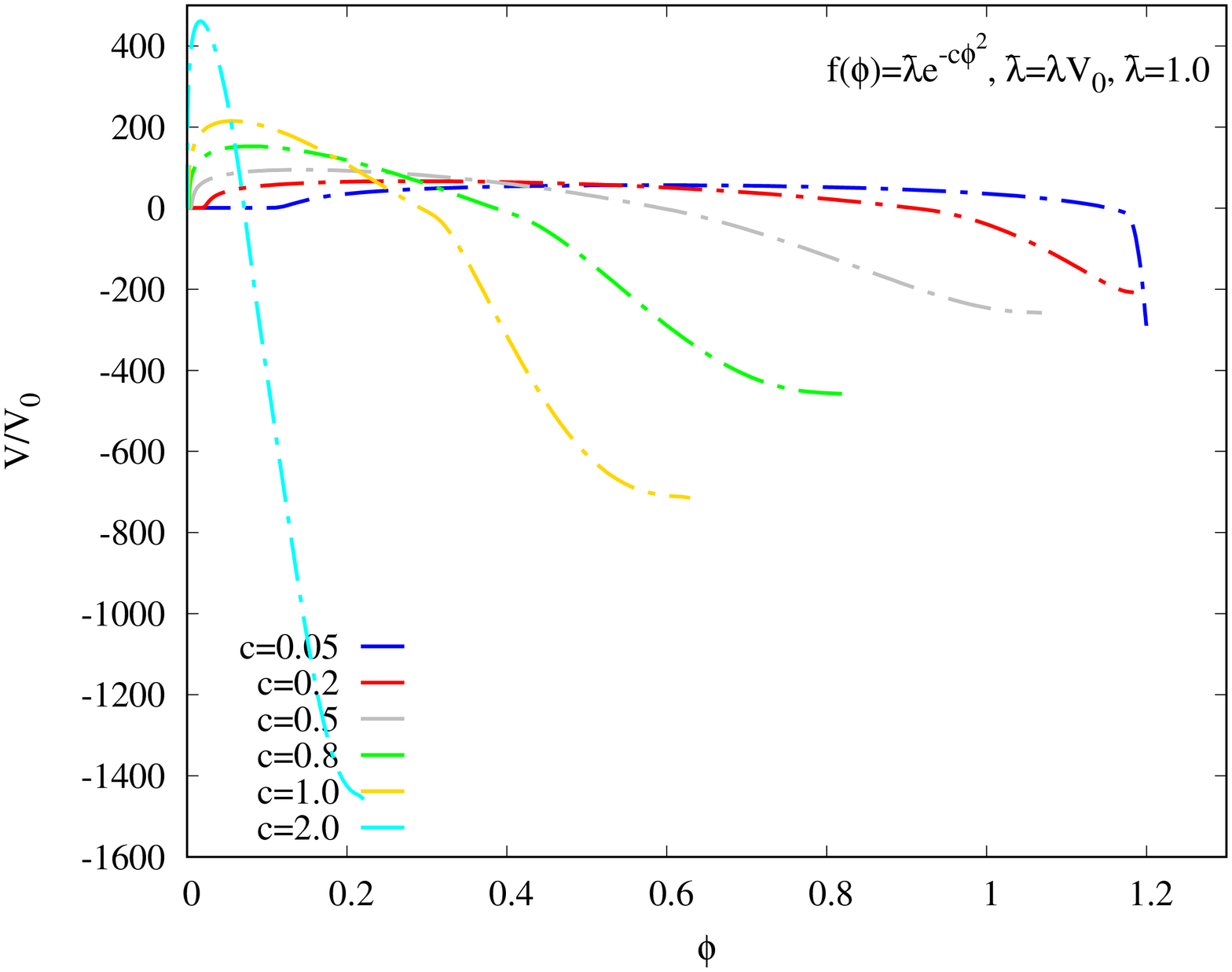}
\includegraphics[scale=0.3,angle=-90]{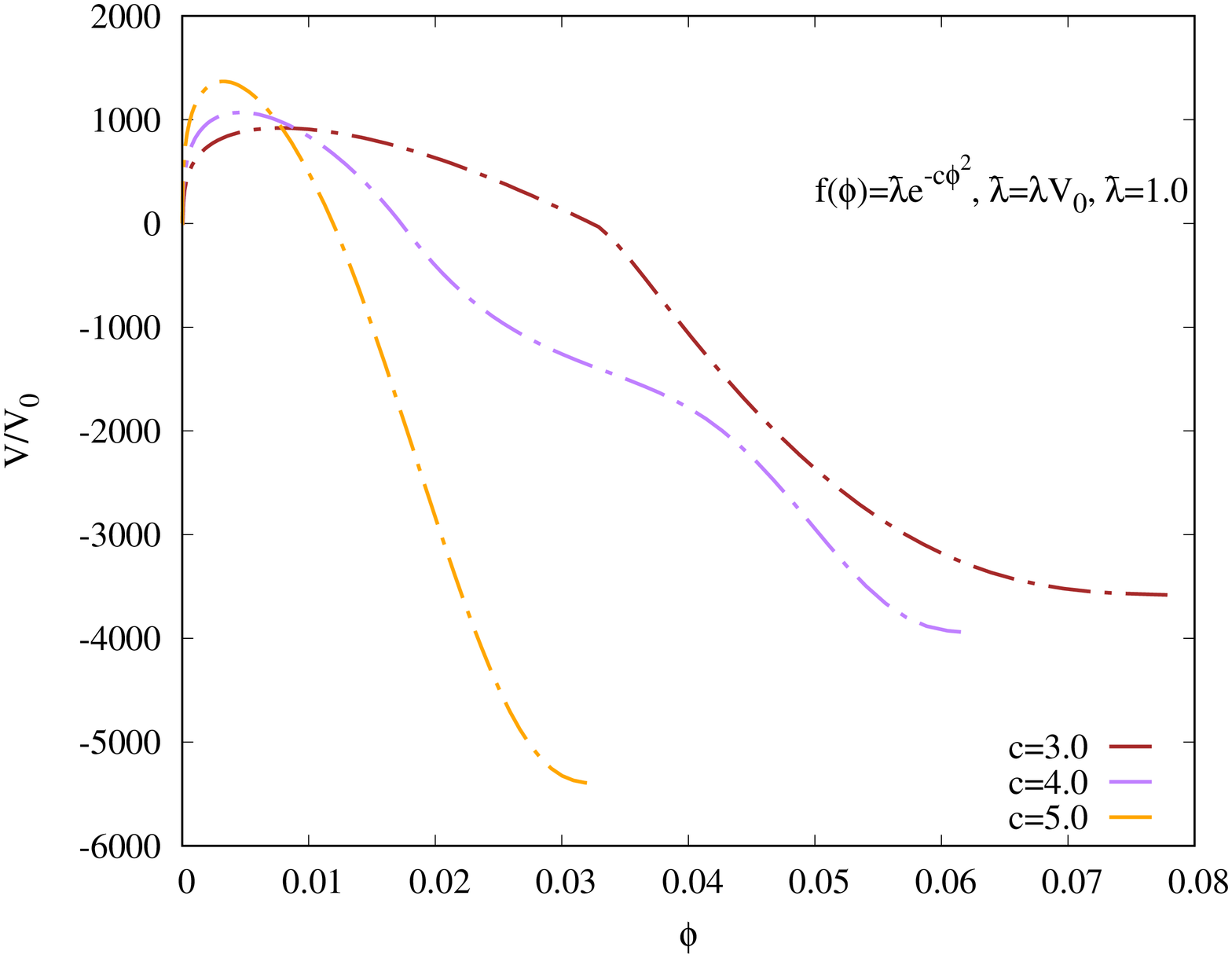}
\caption{\label{fig:lambda2}$a/\sqrt{V_{0}}$ (upper left) and $\dot{a}$ (upper right) versus $\tau/\sqrt{V_{0}}$, and $V(\phi)/V_{0}$ (lower left and right) for $\xi_{2}$, where $\lambda V_{0} = 1$, $\phi_{0} = 0$, $\Delta = 0.8$, and varying $c$; to satisfy the boundary condition, we need to tune $\phi_{1}$ and $V_{0}$.}
\end{center}
\end{figure}

\section{\label{sec:res}Results}

\subsection{Various solutions}

Figs.~\ref{fig:cc1} and \ref{fig:lambda1} are the solutions of the $\xi_{1}(\phi)$ model. In Fig.~\ref{fig:cc1}, we fixed $c = 1$, $\phi_{0} = 0$, $\Delta = 0.8$, and varied $\lambda$; in Fig.~\ref{fig:lambda1}, we fixed $\lambda = 1$, $\phi_{0} = 0$, $\Delta = 0.8$, and varied $c$. For both cases, to satisfy the boundary condition, we tuned $V_{0}$ and $\phi_{1}$; however, we plotted the $V_{0}$-independent result because $V_{0}$ can be rescaled. Therefore, the only physical tuning parameter is reduced to $\phi_{1}$.

Figs.~\ref{fig:cc2} and \ref{fig:lambda2} show the solutions of the $\xi_{2}(\phi)$ model. In Fig.~\ref{fig:cc2}, we fixed $c = 1$, $\phi_{0} = 0$, $\Delta = 0.8$, and varied $\lambda$; in Fig.~\ref{fig:lambda2}, we fixed $\lambda = 1$, $\phi_{0} = 0$, $\Delta = 0.8$, and varied $c$. For both cases, to satisfy the boundary condition, we tuned $V_{0}$ and $\phi_{1}$, but we plotted $V_{0}$-independent result because $V_{0}$ can be rescaled. Therefore, the only physical tuning parameter is reduced to  $\phi_{1}$. It is clear that the physical properties are qualitatively the same as those of the $\xi_{1}$ model.

The metric $a(\tau)$ varies from the left end to the right end, while there exists a throat of the Euclidean wormhole, e.g., a point such that $\dot{a} = 0$ and $\ddot{a} > 0$. After the Wick-rotation, from the result of $V(\phi)$, one can notice that the end point is in de Sitter regime, while the field covers over the potential barrier.

\begin{figure}
\begin{center}
\includegraphics[scale=0.3,angle=-90]{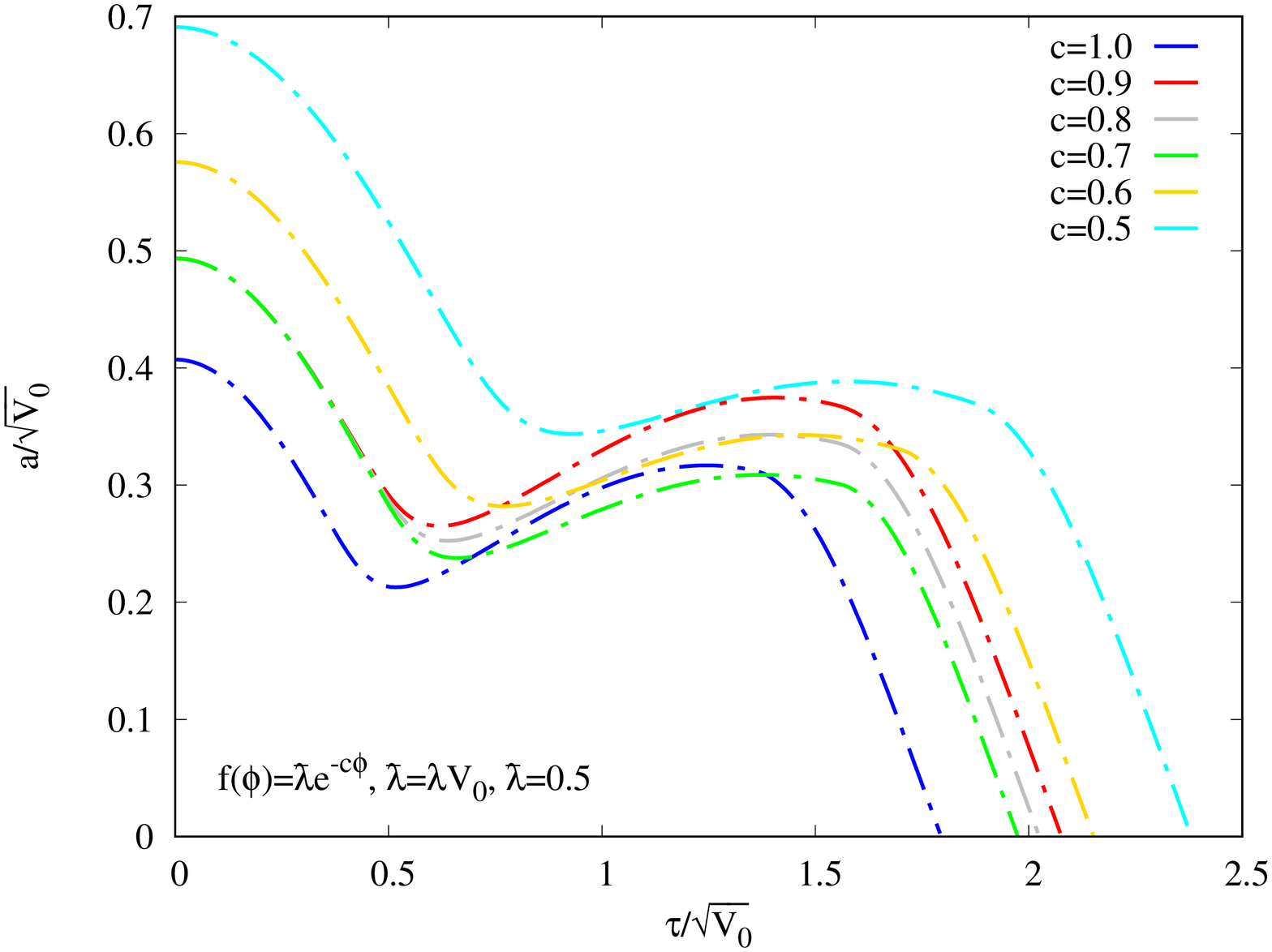}
\includegraphics[scale=0.3,angle=-90]{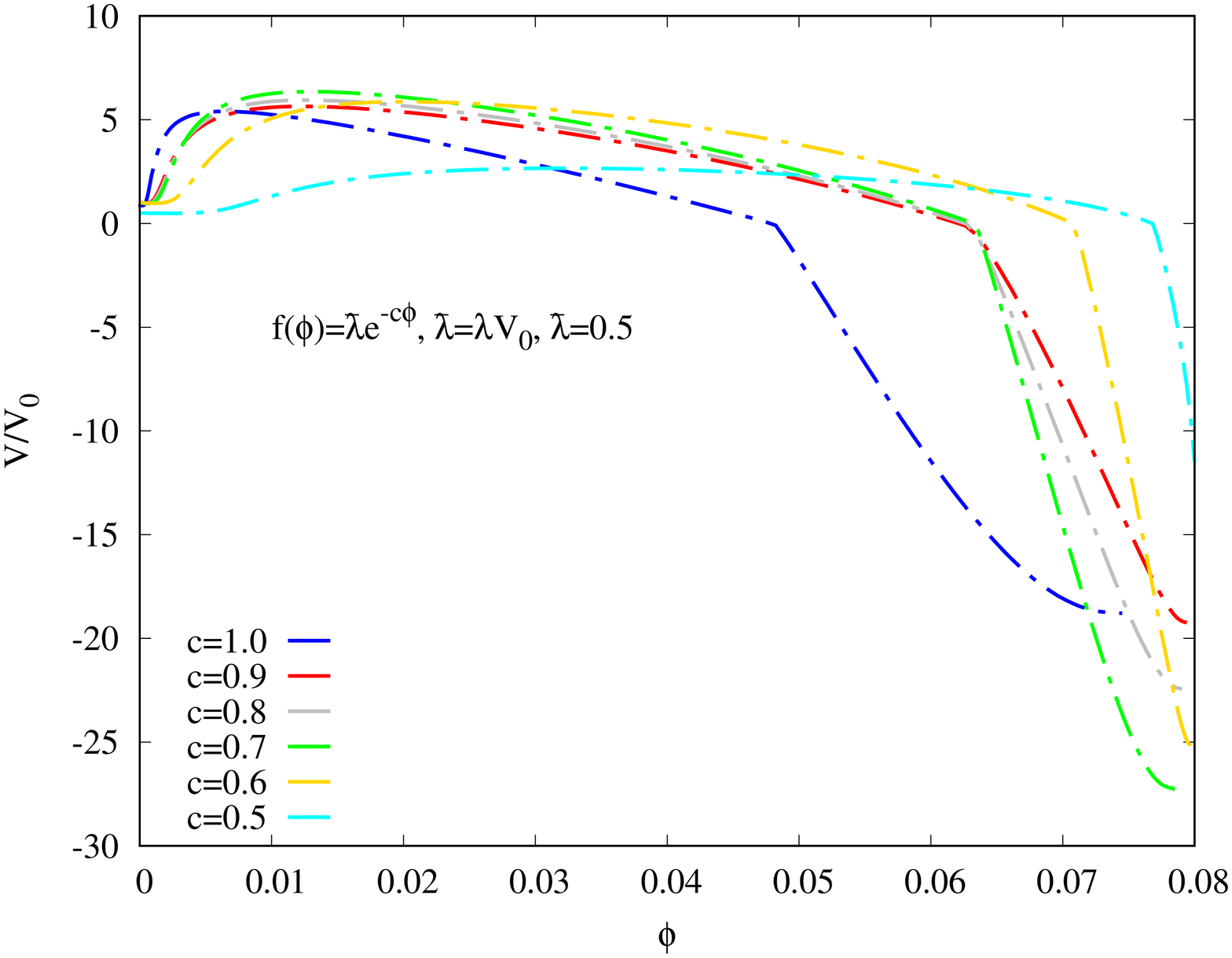}
\caption{\label{fig:subP}$\lambda = 0.5$ with model $\xi_{1}$. The parameters can be on the order of the Planck scale.}
\end{center}
\end{figure}

Of note, when $V_{0}$ is the Planck scale ($\lesssim \mathcal{O}(1)$), in the small $\lambda$ and small $c$ limit, one will find (sub-)Planckian Euclidean wormhole solutions, i.e., all parameter spaces ($V$, $\phi$, etc.) are approximately Planckian, at least for the model $\xi_{1}(\phi)$ (string-inspired model, Fig.~\ref{fig:subP}).

However, for large $\lambda$ or $c$ limit, we still have solutions, but we have some comments on their meaning. One may consider scaling to observe their dependence. Regarding scaling, there is an interplay between $V$ and $\xi$; up to $V_{0}$, it scales by $V/V_{0}$ and $\xi V_{0}$. For a given Euclidean wormhole solution, with an increase in $V_{0}$, the physical potential $V/V_{0}$ becomes smaller; however, at the same time, the physical value of the coupling $\xi$ must increase accordingly. This implies that there is an interplay between two sectors; either the potential is dominant or the Gauss-Bonnet term is dominant, although it describes the same solution. It is fair to say that with an increase in $V_{0}$, the Gauss-Bonnet term is dominant; this may indicate that one needs to include higher order stringy corrections at the same time; however, if $V_{0}$ decreases, the potential term is dominant, which indicates that higher order corrections to the dilaton field must be considered. Therefore, one can conclude that the physical importance of such super-Planckian Gauss-Bonnet-dilaton wormhole is genuinely the non-perturbative quantum gravitational issue. Future investigations on quantum gravity will eventually clarify the true existence of Euclidean wormholes.

\begin{figure}
\begin{center}
\includegraphics[scale=0.5]{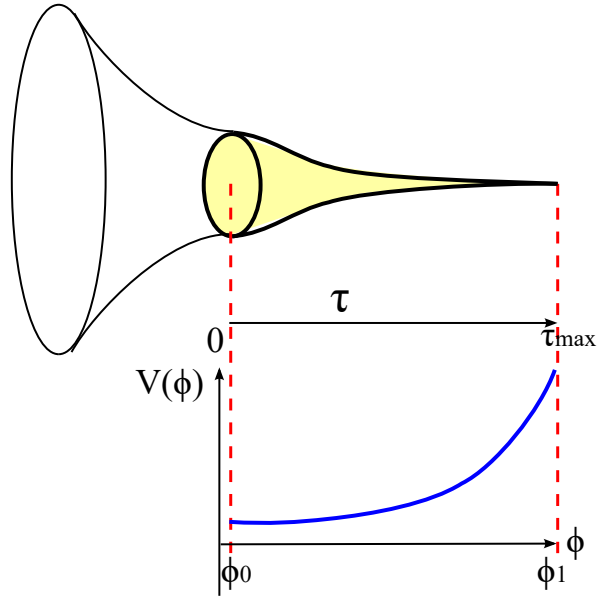}
\caption{\label{fig:fig4}Special form of the solution when $c \ll 1$.}
\end{center}
\end{figure}

One exceptional limit is the blue colored curves in Figs.~\ref{fig:cc1} and \ref{fig:lambda1}, where the final condition satisfies $a(\tau_{\mathrm{max}}) = \dot{a}(\tau_{\mathrm{max}}) = 0$ (Fig.~\ref{fig:fig4}). As long as the Euclidean manifold is located in the quantum regime, there is no reason to disallow such possibility. For this case, the corresponding potential has a different physical property; the point $\tau_{\mathrm{max}}$ corresponds to a certain point of the potential which has a bigger vacuum energy than the left end. These solutions can be obtained if $c \ll 1$. This means that this solution can be free from the scale issue. By choosing proper $V_{0}$, one can find a model so that both the potential and dilaton coupling are the sub-Planckian region. We left the physical importance of this solution as a future topic. Of note, this solution is qualitatively similar to instantons motivated from the loop quantum cosmology \cite{Brahma:2018elv}.

\subsection{Probability}

The probability of each instanton can be evaluated as follows:
\begin{eqnarray}
P \propto e^{-2 B},
\end{eqnarray}
where
\begin{eqnarray}
B = \int_{0}^{\tau_{\mathrm{max}}} L_{E} d\tau.
\end{eqnarray}
For the typical case, if the field is located at the left local minimum of the potential ($\phi = \phi_{0}$), the probability is the same as the Hawking-Moss instantons, i.e.,
\begin{eqnarray}
2B = - \frac{3}{8V_{0}}.
\end{eqnarray}
Note that the value of the Euclidean action is negative definite; hence, with an increase in the volume of instanton, its probability increases. Therefore, it is easy to imagine that the probabilities of instantons, including Euclidean wormholes, are higher than those of the Hawking-Moss instantons.

\begin{figure}
\begin{center}
\includegraphics[scale=0.3,angle=-90]{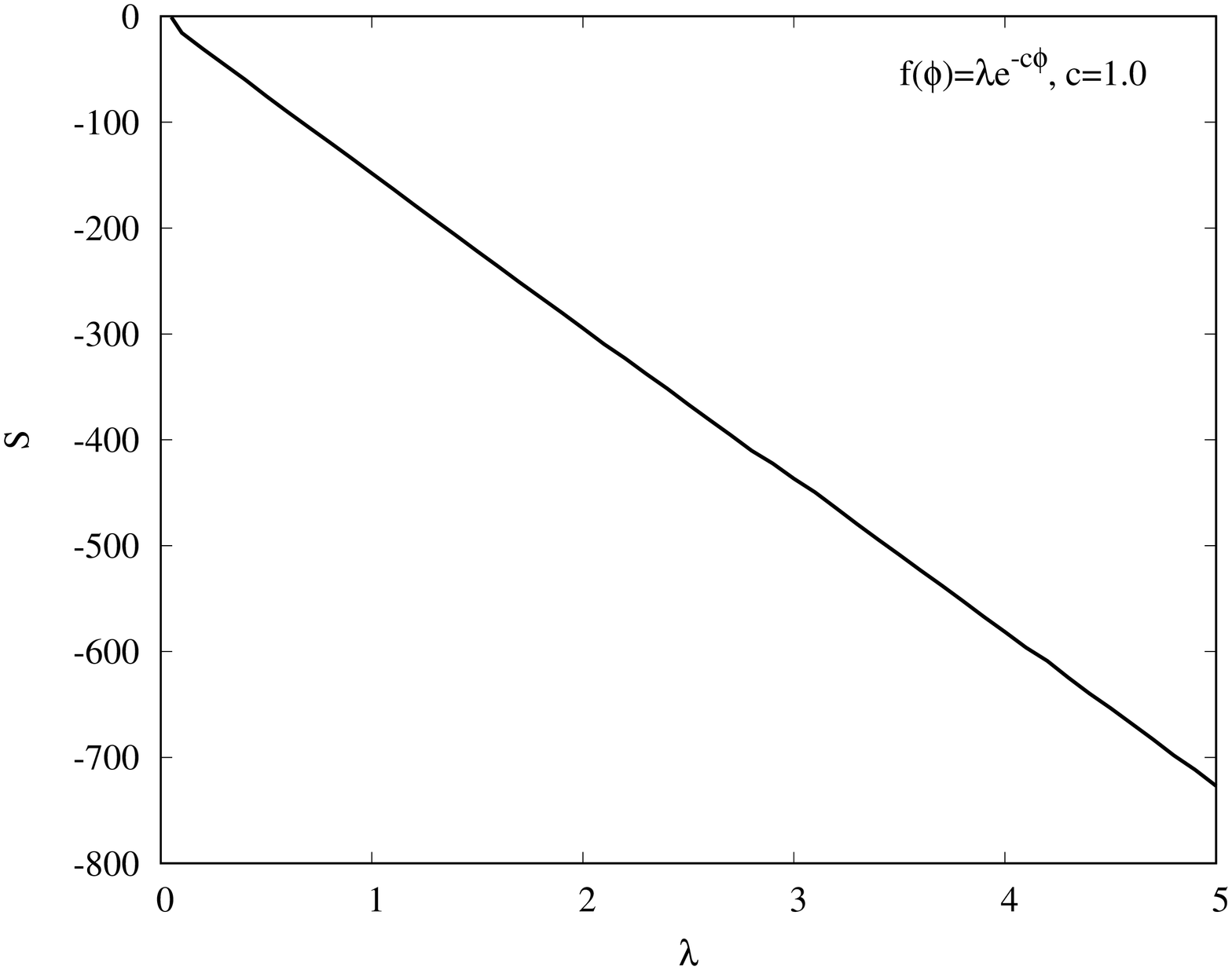}
\includegraphics[scale=0.3,angle=-90]{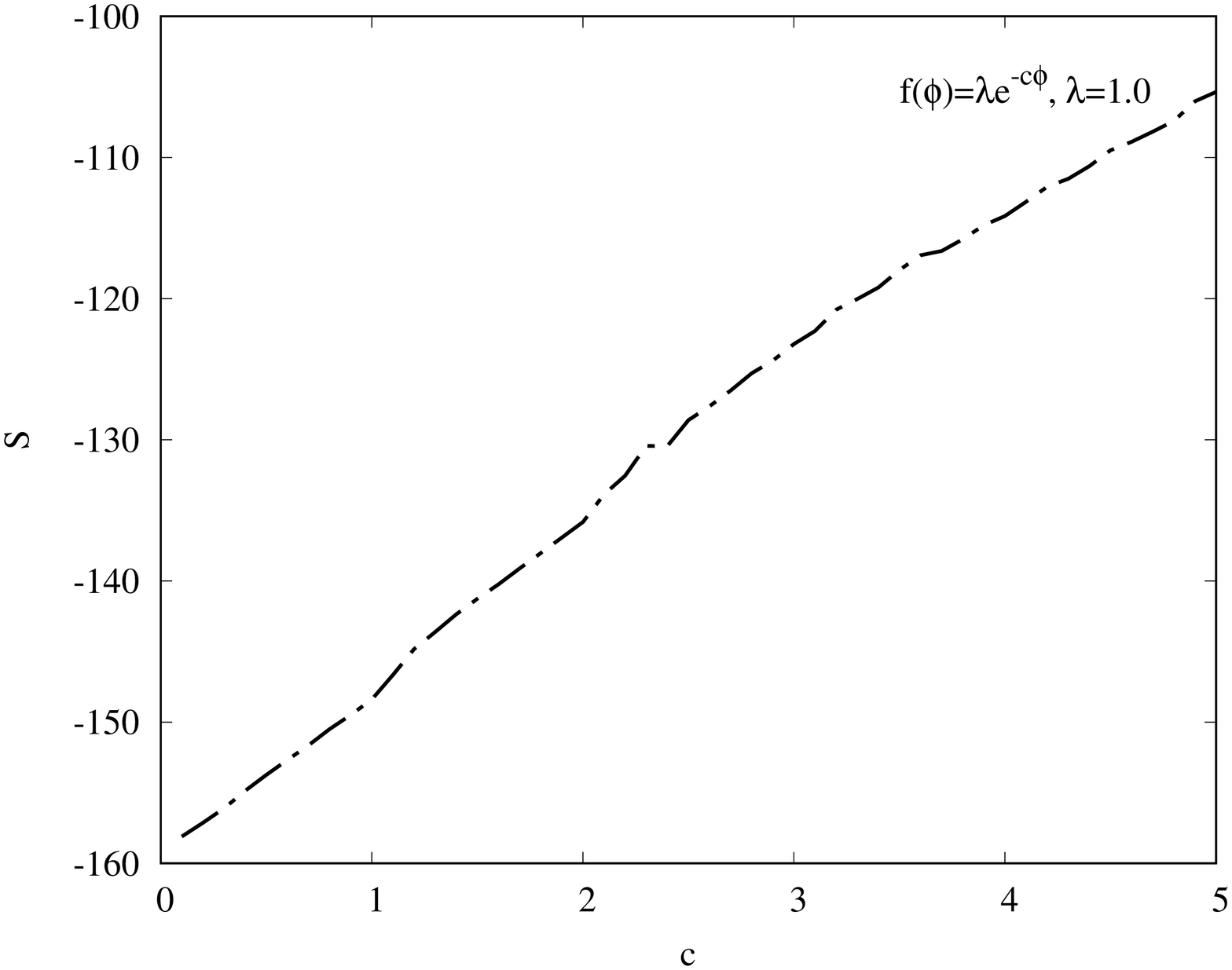}
\caption{\label{fig:Sint}Euclidean action integral ($B$) by varying $\lambda$ (left) and $c$ (right) for model $\xi_{1}(\phi)$.}
\end{center}
\end{figure}

Fig.~\ref{fig:Sint} shows the typical behavior of factor $B$ by varying $\lambda$ and $c$. With an increase in $\lambda$, the Euclidean action becomes more negative; thus, the probability increases. However, with an increase in $c$, the Euclidean action increases; thus, large $c$ limit is less preferred. Of note, except for several parameter spaces, Euclidean wormholes are preferred compared to Hawking-Moss type instantons. Therefore, one may further conclude that Euclidean wormholes can appear even though the slow-roll condition is violated; this instanton can be a new alternative origin of the universe. 

\section{\label{sec:con}Conclusion}

In this study, we investigated Euclidean wormhole solutions in Gauss-Bonnet-dilaton gravity. Specifically, we were interested in the homogeneous analytic continuation to de Sitter space. We observed that the Gauss-Bonnet-dilaton model allowed various types of Euclidean wormhole solutions, and these solutions had higher probability than that of the Hawking-Moss instantons with the same final condition. Therefore, in terms of the probability, Euclidean wormholes are a probable explanation of the origin of our universe, at least in the Gauss-Bonnet-dilaton gravity.

First, although the parameters are limited, by choosing specific parameters, we can determine that the solutions are within the sub-Planckian limit. These solutions have meaningful and important future applications. In addition, if $c \ll 1$, there are no Euclidean wormholes, but a long stretched instanton solution appears. In terms of the homogeneous analytic continuation, this will have no fundamental problem because the stretched end will not be naked to the Lorentzian domain. For these cases, one may find sub-Planckian instanton, which differs from the trivial Hawking-Moss instantons. Of course, its physical importance must be discussed in future investigations.

In addition, there are many parameters that provide super-Planckian Euclidean wormholes. For these solutions, there is an interplay between the Gauss-Bonnet and potential sectors. For a given solution, one can choose a parameter so that the potential term is dominant; or the Gauss-Bonnet term is dominant. For any case, if each term is greater than the Planck scale, it may be theoretically unstable. This issue must be clarified by studying quantum gravity more thoroughly. This is not surprising because we are studying the genuine non-perturbative regime of quantum gravity.

We have several interesting new solutions to describe the origin of our universe. If we turn on perturbations to the manifold, we can evaluate the effect on the anisotropy of the cosmic microwave background \cite{Halliwell:1984eu,Chen:2017aes}. Specifically, the perturbations from the Euclidean wormhole may not result in the scale invariant Bunch-Davies-like state. This result provides interesting observational signatures to cosmological observations; for an ideal case, we expect that cosmological future experiments can confirm or falsify several candidates of quantum gravitational scenarios.

\section*{Acknowledgment}

DY and XYC were supported by the National Research Foundation of Korea (Grant No.: 2018R1D1A1B07049126). GT was supported by IBS under the project code IBS-R018-D1 and by the National Research Foundation of Korea (NRF2016R1D1A1B04932574).


\newpage


\begin{thebibliography}{200}

\bibitem{Metsaev:1987zx}
R.~Metsaev and A.~A.~Tseytlin,
Nucl.\ Phys.\ B \textbf{293}, 385-419 (1987).

\bibitem{Kanti:1995vq}
P.~Kanti, N.~Mavromatos, J.~Rizos, K.~Tamvakis and E.~Winstanley,
Phys.\ Rev.\ D \textbf{54}, 5049-5058 (1996)
[arXiv:hep-th/9511071 [hep-th]].

\bibitem{DeWitt:1967yk}
B.~S.~DeWitt,
Phys.\ Rev.\  \textbf{160}, 1113-1148 (1967).

\bibitem{Hartle:1983ai}
J.~Hartle and S.~Hawking,
Adv.\ Ser.\ Astrophys.\ Cosmol.\  \textbf{3}, 174-189 (1987).

\bibitem{Chen:2018aij}
P.~Chen, M.~Sasaki and D.~Yeom,
Eur.\ Phys.\ J.\ C \textbf{79}, no.7, 627 (2019)
[arXiv:1806.03766 [hep-th]].

\bibitem{Hawking:1981fz}
S.~Hawking and I.~Moss,
Adv.\ Ser.\ Astrophys.\ Cosmol.\  \textbf{3}, 154-157 (1987).

\bibitem{Chen:2016ask}
P.~Chen, Y.~Hu and D.~Yeom,
JCAP \textbf{07}, 001 (2017)
[arXiv:1611.08468 [gr-qc]];\\
S.~Kang and D.~Yeom,
Phys.\ Rev.\ D \textbf{97}, no.12, 124031 (2018)
[arXiv:1703.07746 [gr-qc]];\\
P.~Chen and D.~Yeom,
Eur.\ Phys.\ J.\ C \textbf{78}, no.10, 863 (2018)
[arXiv:1706.07784 [gr-qc]];\\
M.~Bouhmadi-López, C.~Chen, P.~Chen and D.~Yeom,
JCAP \textbf{10}, 056 (2018)
[arXiv:1809.06579 [gr-qc]].

\bibitem{Tumurtushaa:2018agq}
G.~Tumurtushaa and D.~Yeom,
Eur.\ Phys.\ J.\ C \textbf{79}, no.6, 488 (2019)
[arXiv:1808.01103 [hep-th]].

\bibitem{Lee:2012qv}
B.~Lee, W.~Lee and D.~Yeom,
Int.\ J.\ Mod.\ Phys.\ A \textbf{28}, 1350082 (2013)
[arXiv:1206.7040 [hep-th]].


\bibitem{Hwang:2011mp}
D.~Hwang, H.~Sahlmann and D.~Yeom,
Class.\ Quant.\ Grav.\  \textbf{29}, 095005 (2012)
[arXiv:1107.4653 [gr-qc]];\\
D.~Hwang, B.~Lee, H.~Sahlmann and D.~Yeom,
Class.\ Quant.\ Grav.\  \textbf{29}, 175001 (2012)
[arXiv:1203.0112 [gr-qc]];\\
D.~Hwang, S.~A.~Kim, B.~Lee, H.~Sahlmann and D.~Yeom,
Class.\ Quant.\ Grav.\  \textbf{30}, 165016 (2013)
[arXiv:1207.0359 [gr-qc]];\\
D.~Hwang and D.~Yeom,
JCAP \textbf{06}, 007 (2014)
[arXiv:1311.6872 [gr-qc]].

\bibitem{Chen:2019cmw}
P.~Chen, D.~Ro and D.~Yeom,
Phys.\ Dark Univ.\  \textbf{28}, 100492 (2020)
[arXiv:1904.00199 [gr-qc]].

\bibitem{Cai:2008ht}
R.~Cai, B.~Hu and S.~Koh,
Phys.\ Lett.\ B \textbf{671}, 181-186 (2009)
[arXiv:0806.2508 [hep-th]].

\bibitem{Koh:2014bka}
S.~Koh, B.~Lee, W.~Lee and G.~Tumurtushaa,
Phys.\ Rev.\ D \textbf{90}, no.6, 063527 (2014)
[arXiv:1404.6096 [gr-qc]];\\
B.~Lee, W.~Lee and D.~Ro,
Phys.\ Lett.\ B \textbf{762}, 535-542 (2016)
[arXiv:1607.01125 [hep-th]];\\
S.~Koh, B.~Lee and G.~Tumurtushaa,
Phys.\ Rev.\ D \textbf{95}, no.12, 123509 (2017)
[arXiv:1610.04360 [gr-qc]];\\
S.~Koh, B.~Lee and G.~Tumurtushaa,
Phys.\ Rev.\ D \textbf{98}, no.10, 103511 (2018)
[arXiv:1807.04424 [astro-ph.CO]].


\bibitem{Kanno:2012zf}
S.~Kanno, M.~Sasaki and J.~Soda,
Class.\ Quant.\ Grav.\  \textbf{29}, 075010 (2012)
[arXiv:1201.2272 [hep-th]];\\
S.~Kanno, M.~Sasaki and J.~Soda,
Prog.\ Theor.\ Phys.\  \textbf{128}, 213-226 (2012)
[arXiv:1203.0612 [hep-th]].

\bibitem{Brahma:2018elv}
S.~Brahma and D.~Yeom,
Phys.\ Rev.\ D \textbf{98}, no.8, 083537 (2018)
[arXiv:1808.01744 [gr-qc]];\\
S.~Brahma and D.~Yeom,
Universe \textbf{5}, no.1, 22 (2019)
[arXiv:1810.10211 [hep-th]].

\bibitem{Halliwell:1984eu}
J.~Halliwell and S.~Hawking,
Adv.\ Ser.\ Astrophys.\ Cosmol.\  \textbf{3}, 277-291 (1987).

\bibitem{Chen:2017aes}
P.~Chen, Y.~Lin and D.~Yeom,
Eur.\ Phys.\ J.\ C \textbf{78}, no.11, 930 (2018)
[arXiv:1707.01471 [gr-qc]];\\
P.~Chen, H.~Yeh and D.~Yeom,
Phys.\ Dark Univ.\  \textbf{27}, 100435 (2020)
[arXiv:1903.12045 [gr-qc]].


\bibitem{Glavan:2019inb}
D.~Glavan and C.~Lin,
Phys. Rev. Lett. \textbf{124}, no.8, 081301 (2020)
[arXiv:1905.03601 [gr-qc]].

\bibitem{Lu:2020iav}
H.~Lu and Y.~Pang,
[arXiv:2003.11552 [gr-qc]].

\bibitem{Guo:2009uk}
Z.~Guo and D.~J.~Schwarz,
Phys. Rev. D \textbf{80}, 063523 (2009)
[arXiv:0907.0427 [hep-th]].

Z.~Guo and D.~J.~Schwarz,
Phys. Rev. D \textbf{81}, 123520 (2010)
[arXiv:1001.1897 [hep-th]].

P.~Jiang, J.~Hu and Z.~Guo,
Phys. Rev. D \textbf{88}, 123508 (2013)
[arXiv:1310.5579 [hep-th]].

C.~van de Bruck and C.~Longden,
Phys. Rev. D \textbf{93}, no.6, 063519 (2016)
[arXiv:1512.04768 [hep-ph]].

C.~van de Bruck, K.~Dimopoulos and C.~Longden,
Phys. Rev. D \textbf{94}, no.2, 023506 (2016)
[arXiv:1605.06350 [astro-ph.CO]].

I.~Fomin and S.~Chervon,
Grav. Cosmol. \textbf{23}, no.4, 367-374 (2017)
[arXiv:1704.03634 [gr-qc]].

Z.~Yi, Y.~Gong and M.~Sabir,
Phys. Rev. D \textbf{98}, no.8, 083521 (2018)
[arXiv:1804.09116 [gr-qc]].

S.~Chakraborty, T.~Paul and S.~SenGupta,
Phys. Rev. D \textbf{98}, no.8, 083539 (2018)
[arXiv:1804.03004 [gr-qc]].

S.~Odintsov and V.~Oikonomou,
Phys. Rev. D \textbf{98}, no.4, 044039 (2018)
[arXiv:1808.05045 [gr-qc]].


\bibitem{Calcagni:2005im}
G.~Calcagni, S.~Tsujikawa and M.~Sami,
Class. Quant. Grav. \textbf{22}, 3977-4006 (2005)
[arXiv:hep-th/0505193 [hep-th]].

S.~Nojiri, S.~D.~Odintsov and M.~Sami,
Phys. Rev. D \textbf{74}, 046004 (2006)
[arXiv:hep-th/0605039 [hep-th]].

B.~M.~Leith and I.~P.~Neupane,
JCAP \textbf{05}, 019 (2007)
[arXiv:hep-th/0702002 [hep-th]].

G.~Cognola, E.~Elizalde, S.~Nojiri, S.~Odintsov and S.~Zerbini,
Phys. Rev. D \textbf{75}, 086002 (2007)
[arXiv:hep-th/0611198 [hep-th]].

S.~Nojiri, S.~D.~Odintsov and M.~Sasaki,
Phys. Rev. D \textbf{71}, 123509 (2005)
[arXiv:hep-th/0504052 [hep-th]].

S.~Nojiri and S.~D.~Odintsov,
Phys. Lett. B \textbf{631}, 1-6 (2005)
[arXiv:hep-th/0508049 [hep-th]].

S.~Nojiri, S.~D.~Odintsov and P.~V.~Tretyakov,
Phys. Lett. B \textbf{651}, 224-231 (2007)
[arXiv:0704.2520 [hep-th]].

K.~Bamba, A.~N.~Makarenko, A.~N.~Myagky and S.~D.~Odintsov,
JCAP \textbf{04}, 001 (2015)
[arXiv:1411.3852 [hep-th]].

\bibitem{Gong:2017kim}
Y.~Gong, E.~Papantonopoulos and Z.~Yi,
Eur. Phys. J. C \textbf{78}, no.9, 738 (2018)
[arXiv:1711.04102 [gr-qc]].

M.~Heydari-Fard, H.~Razmi and M.~Yousefi,
Int. J. Mod. Phys. D \textbf{26}, no.02, 1750008 (2016)



\end{thebibliography}
\end{document}